\documentclass{aa}  
\pdfoutput=1  
\usepackage{graphicx}
\usepackage{txfonts}
\usepackage{hyperref}
\begin{document}

   \title{Revisiting Andromeda's Parachute\thanks{Tables 2, 4, 6, and 7 are only 
    available in electronic form at the CDS via anonymous ftp to cdsarc.u-strasbg.fr 
    (130.79.128.5) or via http://cdsweb.u-strasbg.fr/cgi-bin/qcat?J/A+A/vol/page}}    
                                                             
   \author{Vyacheslav N. Shalyapin\inst{1,2,3}
                \and
                Luis J. Goicoechea\inst{1}
                \and
                Karianne Dyrland\inst{4,5}
                \and
                H\r{a}kon Dahle\inst{4}       
                }

\institute{Departamento de F\'\i sica Moderna, Universidad de Cantabria, 
                Avda. de Los Castros s/n, E-39005 Santander, Spain\\
                \email{vshal@ukr.net;goicol@unican.es}
                \and
                O.Ya. Usikov Institute for Radiophysics and Electronics, National 
                Academy of Sciences of Ukraine, 12 Acad. Proscury St., UA-61085 
                Kharkiv, Ukraine
                \and
                Institute of Astronomy of V.N. Karazin Kharkiv National University,
                Svobody Sq. 4, UA-61022 Kharkiv, Ukraine
                \and
		Institute of Theoretical Astrophysics, University of Oslo, PO Box 
                1029, Blindern 0315, Oslo, Norway
		\and
                Kongsberg Defence \& Aerospace AS, Instituttveien 10, PO Box 26, 
                2027, Kjeller, Norway}


 
  \abstract{The gravitational lens system PS J0147+4630 (Andromeda's Parachute) consists 
of four quasar images ABCD and a lensing galaxy. We obtained $r$-band light curves of ABCD 
in the 2017$-$2021 period from a monitoring with two 2-m class telescopes. These curves and 
state-of-the-art curve-shifting algorithms led to three independent time delays  
relative to image A, one of which is accurate enough (uncertainty of about 4\%) to be used 
in cosmological studies. Our finely sampled light curves and some additional fluxes in the 
years 2010$-$2013 also demonstrated the presence of significant microlensing variations. 
This paper also focused on new near-IR spectra of ABCD in 2018$-$2019 that were derived from 
archive data of two 10-m class telescopes. We analysed the spectral region including the 
Mg\,{\sc ii}, H$\beta$, [O\,{\sc iii}], and H$\alpha$ emission lines (0.9$-$2.4 $\mu$m), 
measuring image flux ratios and a reliable quasar redshift of 2.357 $\pm$ 0.002, and finding 
evidence of an outflow in the H$\alpha$ emission. In addition, we updated the lens mass 
model of the system and estimated a quasar black-hole logarithmic mass ${\log 
\left[ M_{\rm{BH}}/\rm{M_{\odot}} \right]}$ = 9.34 $\pm$ 0.30.} 
    
   \keywords{gravitational lensing: strong -- 
                quasars: individual: PS J0147+4630}

   \maketitle
%

\section{Introduction}
\label{sec:intro}

\begin{figure*}
\centering
\includegraphics[width=14cm]{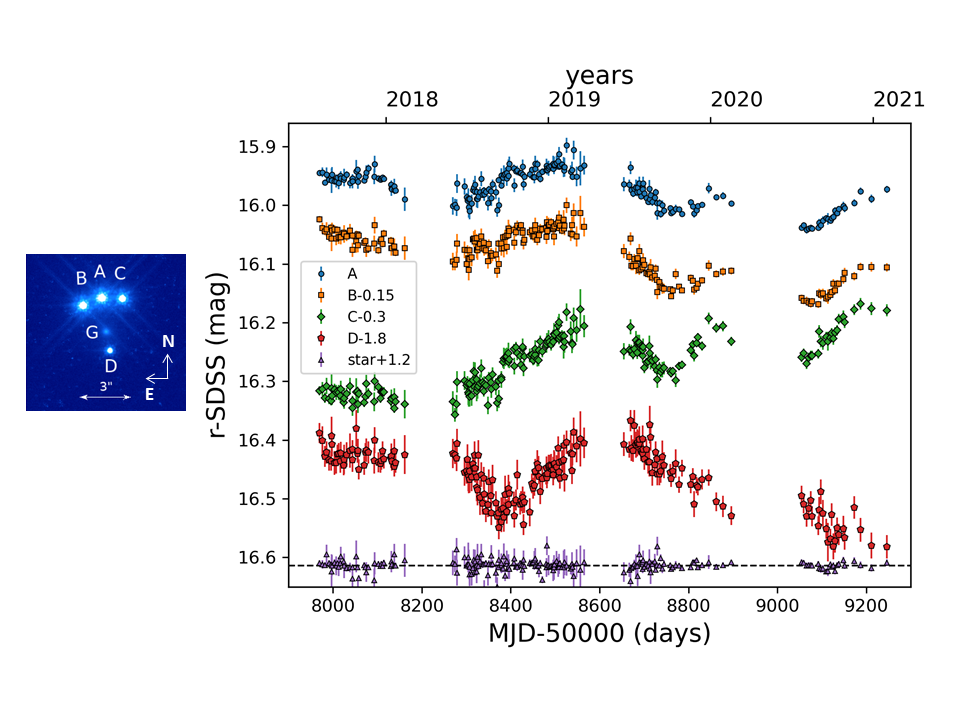}
\caption{{\it Left}: quasar images ABCD and main lens galaxy G of PS J0147+4630 from a 
public HST-WFC3 frame of the system in the F814W band. {\it Right}: LT-NOT light curves of 
PS J0147+4630 from its discovery to 2021. The $r$-band magnitudes of images B, C, and D, and 
the control star are offset by $-$0.15, $-$0.3, $-$1.8, and +1.2, respectively, to 
facilitate comparison between them and with image A.}
\label{fig:f1}
\end{figure*}

Optical frames from the Panoramic Survey Telescope and Rapid Response System 
\citep[Pan-STARRS;][]{2016arXiv161205560C} led to the serendipitous discovery of the strong 
gravitational lens system with a quadruply-imaged quasar (quad) \object{PS J0147+4630} 
\citep{2017ApJ...844...90B}. Due to its position in the sky and the spatial arrangement of 
the four quasar images, this quad is also called Andromeda's Parachute 
\citep[e.g.][]{2018ApJ...859..146R}. The three brightest images (A, B and C) forms an arc 
that is about 3\arcsec\ from the faintest image D, and the main lens galaxy G is located 
between the bright arc and D. This configuration is clearly seen in the left panel of 
Figure~\ref{fig:f1}, which is based on Hubble Space Telescope (HST) data. 

As far as we know, the quasar \object{PS J0147+4630} is the brightest source in the sky at 
redshifts $z$ > 1.4 (apart from transient events such as gamma-ray bursts), and its four 
optical images can be easily resolved with a ground-based telescope in normal seeing 
conditions. Thus, it is a compelling target for various physical studies based on 
high-resolution spectroscopy \citep[e.g.][]{2018ApJ...859..146R} and detailed photometric 
monitoring \citep[e.g.][]{2018MNRAS.475.3086L}. Early two-season monitoring 
campaigns with the 2.0 m Liverpool Telescope \citep[LT;][]{2019ApJ...887..126G} and the 2.5 
m Nordic Optical Telescope \citep[NOT;][]{MScKD} provided accurate optical light curves of 
all quasar images, as well as preliminary time delays and evidence of microlensing-induced 
variations. A deeper look at the optical variability of Andromeda's Parachute is of 
great importance, since robust time delays and well-observed microlensing variations can be 
used to determine cosmological parameters \citep[e.g.][]{2016A&ARv..24...11T} and the 
structure of the quasar accretion disc \citep[e.g.][]{2010GReGr..42.2127S}. 

Early optical spectra of the system confirmed the gravitational lensing phenomenon and 
revealed the broad absorption-line (BAL) nature of the quasar \citep{2017A&A...605L...8L,
2018ApJ...859..146R}. However, there is an appreciable discrepancy between the quasar 
redshift reported by \citet{2018ApJ...859..146R} and that measured by 
\citet{2017A&A...605L...8L}, amounting to $\Delta z_{\rm{s}} \sim$ 0.04. 
\citet{2018MNRAS.475.3086L} also performed the first attempt to determine the redshift of G 
from spectroscopic observations with the 8.1 m Gemini North Telescope (GNT). An accurate 
reanalysis of these GNT data showed that the first estimate of the lens redshift was biased, 
by enabling better identification of G as an early-type galaxy at $z_{\rm{l}}$ = 0.678 $\pm$ 
0.001 with stellar velocity dispersion $\sigma_{\rm{l}}$ = 313 $\pm$ 14 km s$^{-1}$ 
\citep{2019ApJ...887..126G}. Both redshifts, $z_{\rm{s}}$ and $z_{\rm{l}}$, are key 
pieces of information to interpret, among other things, time delays and microlensing 
effects. Additionally, HST high-resolution imaging of the lens system provided a lens mass 
model \citep{2019MNRAS.483.5649S,2021MNRAS.501.2833S}. To ensure proper interpretation 
of delays and microlensing-induced phenomena, a reliable lens mass model is also required.   

New spectroscopic observations in the unexplored near-IR region are also useful tools 
to shed light on physical properties of the quad. In addition to the determination of 
redshifts, wavelength-domain data are often used to measure image flux ratios for emission 
lines and their underlying continua. These measurements provide insights about the macrolens 
flux ratios and extinction/microlensing effects \citep[e.g.][]{2012ApJ...755...82M,
2016A&A...596A..77G,2017ApJ...836...14S}. The macrolens flux ratios, for instance, put 
constraints on the distribution of mass lensing the quasar. Moreover, if one has information 
on the magnification and transmission factors for a given quasar image, line widths and 
continuum fluxes for such image yield estimates of the quasar black hole mass 
\citep[e.g.][]{2006ApJ...641..689V,2011ApJ...742...93A}, thus constraining the size of the 
innermost region in the accretion disc.

\begin{table*}
\centering
\caption{Pan-STARRS positions and magnitudes of relevant field stars.}
\begin{tabular}{lccccc}
\hline\hline
Star & RA(J2000) & Dec(J2000) & $g$ & $r$ & $i$ \\
\hline
PSF 	& 26.773246 & 46.506670 & 16.366 & 15.606 & 15.260 \\
Control & 26.746290 & 46.504028 & 15.800 & 15.421 & 15.269 \\
Cal1    & 26.805695 & 46.522834 & 16.587 & 16.292 & 16.208 \\
Cal2    & 26.725610 & 46.488113 & 16.857 & 16.405 & 16.257 \\
Cal3    & 26.752831 & 46.518659 & 17.157 & 16.836 & 16.718 \\
Cal4    & 26.760809 & 46.474513 & 17.229 & 16.856 & 16.714 \\
Cal5    & 26.824027 & 46.528718 & 15.656 & 15.200 & 15.029 \\
Cal6    & 26.790480 & 46.502241 & 15.145 & 14.831 & 14.716 \\
\hline
\end{tabular}
\tablefoot{
Astrometric and photometric data of the stars that we used for PSF fitting (PSF), 
variability checking (Control), and calibration (Cal1-Cal6). RA(J2000) and Dec(J2000) are 
given in degrees.
} 
\label{tab:t1}
\end{table*}

This paper is organized as follows. In Sect.~\ref{sec:lcs}, we present combined LT and NOT 
light curves of the four images of \object{PS J0147+4630} spanning four observing seasons 
from 2017 to 2021. In Sect.~\ref{sec:delmic}, using these optical light curves, we carefully 
analyse the time delays between images and the quasar microlensing variability. In 
Sect.~\ref{sec:spec}, we present an analysis of near-IR spectroscopic data of the 
system in 2018$-$2019, focusing on the quasar redshift and the image flux ratios. In 
Sect.~\ref{sec:mass}, assuming a flat $\Lambda$CDM (standard) cosmology, we discuss the 
Hubble constant ($H_0$) value that is inferred from the lens mass model based on HST 
imaging, updated redshifts, and the longest time delay that we measure. A lens mass 
modelling based on astrometric and time-delay constraints is also discussed in 
Sect.~\ref{sec:mass}. In Sect.~\ref{sec:bhmass}, we measure the mass of the central black 
hole in the quasar. Our main conclusions are included in Sect.~\ref{sec:end}.    

\section{New optical light curves}
\label{sec:lcs}

We monitored \object{PS J0147+4630} with the LT from 2017 August to 2021 February using the 
IO:O optical camera with a pixel scale of $\sim$0\farcs30. Each observing night, a single 
120 s exposure was taken in the Sloan $r$-band filter, and over the full monitoring period, 
145 $r$-band frames were obtained. The LT data reduction pipeline carried out three basic 
tasks: bias subtraction, overscan trimming, and flat fielding. Additionally, the IRAF 
software\footnote{https://iraf-community.github.io/} \citep{1986SPIE..627..733T,
1993ASPC...52..173T} allowed us to remove cosmic rays and bad pixels from all frames. We 
extracted the brightness of the four quasar images ABCD through PSF fitting, using the 
IMFITFITS software \citep{1998AJ....115.1377M} and following the scheme described by 
\citet{2019ApJ...887..126G}. Table~\ref{tab:t1} includes the position and magnitudes of the 
PSF star, as well as of other relevant field stars. These data are taken from the Data 
Release 1 of Pan-STARRS\footnote{http://panstarrs.stsci.edu} 
\citep{2020ApJS..251....7F}. Our photometric model consisted of four point-like sources 
(ABCD) and a de Vaucouleurs profile convolved with the empirical PSF (lensing galaxy G). 
Positions of components with respect to A and structure parameters of G were constrained 
from HST data \citep{2019MNRAS.483.5649S,2021MNRAS.501.2833S}.

We also selected six non-variable blue stars in the field of \object{PS J0147+4630} and 
performed PSF photometry on five of them (see the calibration stars Cal1-Cal5 in 
Table~\ref{tab:t1}; Cal6 is a saturated star in LT frames). For each of the five calibration 
stars, we calculated its average magnitude within the monitoring period and magnitude 
deviations in individual frames (by subtracting average). In each individual frame, the 
five stellar magnitude deviations were averaged together to calculate a single magnitude 
offset, which was then subtracted from the magnitudes of quasar images. After this 
photometric calibration, we removed 15 observing epochs in which quasar magnitudes deviate 
appreciably from adjacent values. Thus, the final LT $r$-band light curves are based on 130 
frames (epochs), and the typical uncertainties in the light curves of the quasar images and 
control star (see Table~\ref{tab:t1}) were estimated from magnitude differences between 
adjacent epochs separated by no more than 4 d \citep{2019ApJ...887..126G}. We derived 
typical errors of 0.0058 (A), 0.0069 (B), 0.0091 (C), 0.0188 (D), and 0.0054 (control 
star) mag. For the control star, we have also verified that its typical error practically 
coincides with the standard deviation of all measures (0.0051 mag). To obtain photometric 
uncertainties at each observing epoch, the typical errors were scaled by the relative 
signal-to-noise ratio of the PSF star \citep{2000hccd.book.....H}.    

The optical monitoring of \object{PS J0147+4630} with the NOT spanned from 2017 August to 
2019 December. We used the ALFOSC camera with a pixel scale of $\sim$0\farcs21 and the 
$R$-Bessel filter. This passband is slightly redder than the Sloan $r$ band. Each observing 
night, we mainly took three exposures of 30 s each under good seeing conditions. The 
full-width at half-maximum (FWHM) of the seeing disc was about 1\arcsec, and we collected 
298 individual frames over the entire monitoring campaign. After a standard data reduction, 
IMFITFITS PSF photometry yielded magnitudes for the quasar images (see above for details on 
the photometric model). To avoid biases in the combined LT-NOT light curves, the same 
photometric method was applied to LT and NOT frames. This method differs from that of 
\citet{MScKD}, who used the DAOPHOT package in IRAF \citep{1987PASP...99..191S,TRMD} to 
extract magnitudes from NOT frames. 

The six calibration stars in Table~\ref{tab:t1} were used to adequately correct quasar 
magnitudes (see above), and we were forced to remove 17 individual frames leading to 
magnitude outliers. We then combined $R$-band magnitudes measured on the same night to 
obtain photometric data of the lensed quasar and control star at 77 epochs. Again, typical 
errors were derived from magnitudes at adjacent epochs that are separated $<$ 4.5 d. This 
procedure led to uncertainties of 0.0122 (A), 0.0122 (B), 0.0144 (C), 0.0197 (D), and 0.0170 
(control star) mag. Errors at each observing epoch were calculated in the same way as for 
the LT light curves. 

As a last step, we combined the $r$-band LT and $R$-band NOT light curves. If we focus on 
the quasar images and consider $rR$ pairs separated by no more than 2.5 d, the values of the 
average colour $\langle r - R \rangle$ are 0.0568 (A), 0.0619 (B), 0.0549 (C), and 0.0655 
(D). Brightness records of the ABC images are more accurate than those of D, and thus we 
reasonably take the average colours of ABC to estimate a mean $r - R$ offset of 0.0579 mag. 
After correcting the $R$-band curves of the quasar for this offset, we obtain the new 
records in Table 2. This machine-readable ASCII file at the CDS contains $r$-band magnitudes 
of the quasar images and the control star at 207 observing epochs (MJD$-$50\,000). In 
Figure~\ref{fig:f1}, we also display our new 4-year light curves.
    
\section{Time delays and microlensing signals}
\label{sec:delmic}

\begin{figure}
\centering
\includegraphics[width=9cm]{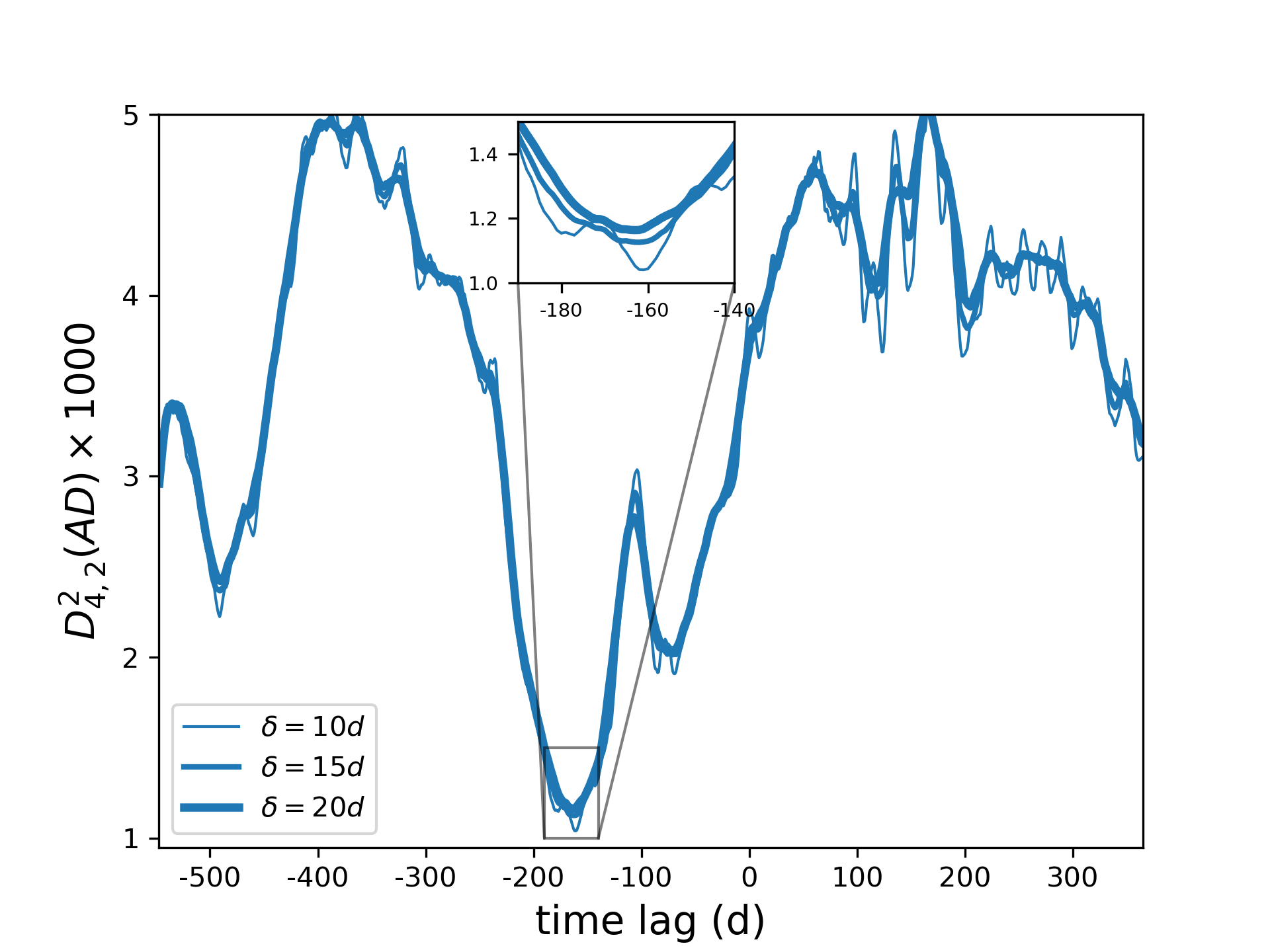}
\caption{$D^2_{4,2}$(AD) dispersion spectra for a SMO model, using decorrelation lengths 
$\delta$ = 10, 15 and 20 d. Each dispersion spectrum describes the values of $D^2_{4,2}$(AD) 
for a broad range of time lags after optimising the magnitude offsets for every lag. We also 
zoom into the region that includes the minima (box in the right top corner).}
\label{fig:f2}
\end{figure}

\begin{figure*}
\centering
\includegraphics[width=14cm]{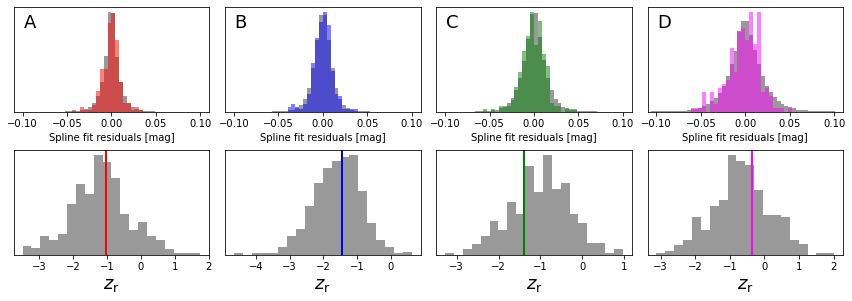}
\caption{{\it Top}: distributions of FKS fit residuals for $\eta$ = 50 d and 
$\eta_{\rm{ml}}$ = 400 d. The grey histograms represent the distributions of residuals from 
the fits to 500 synthetic light curves of each image, while the red, blue, green and magenta 
histograms correspond to the distributions of residuals from the fits to the LT-NOT light 
curves. {\it Bottom}: normalised number of runs $Z_{\rm{r}}$ for the synthetic data (grey 
histograms) and the observed brightness records (red, blue, green and magenta vertical 
lines).}
\label{fig:f3}
\end{figure*} 

\begin{figure*}
\centering
\includegraphics[width=14cm]{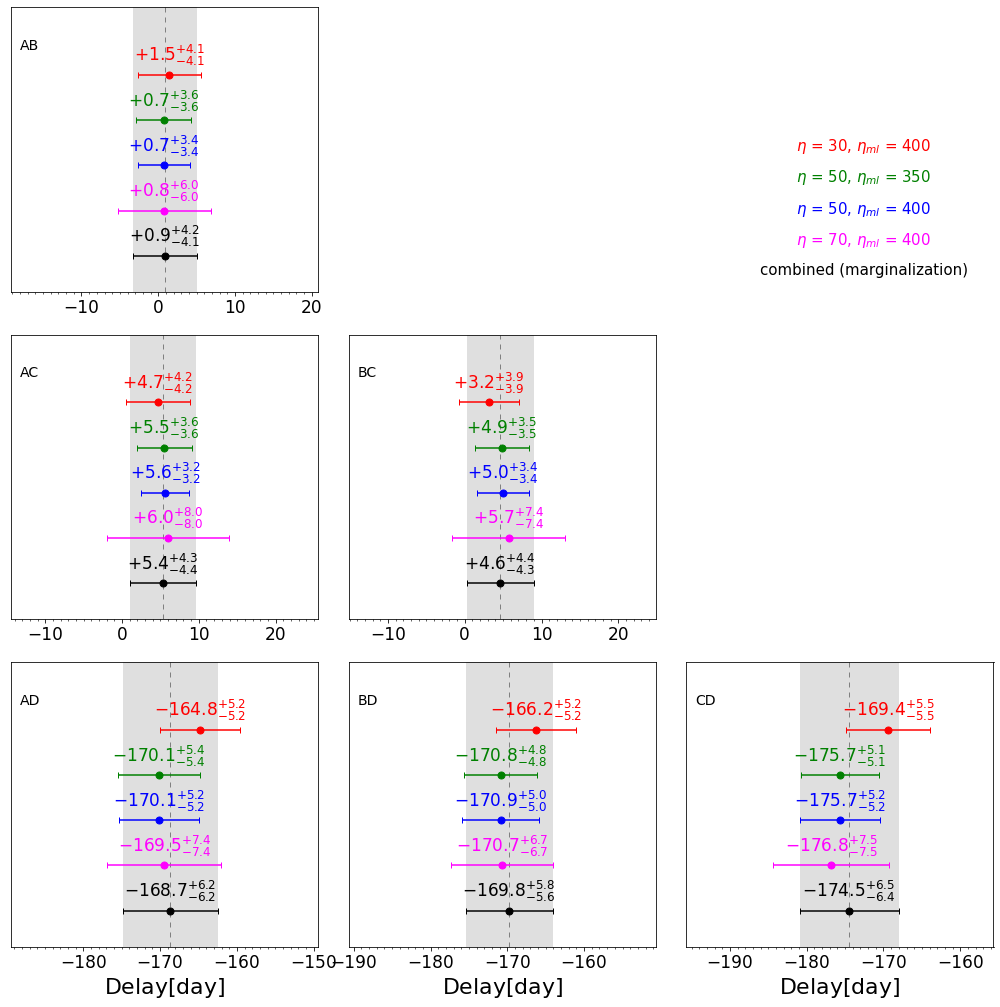}
\caption{Time-delay estimates using the $\chi^2$ technique and free-knot splines. Combined 
estimates \citep[$\tau_{\rm{thresh}}$ = 0;][]{2018A&A...616A.183B} are highlighted with grey 
rectangles encompassing the individual measurements.}
\label{fig:f4}
\end{figure*} 

Previous efforts focused on early monitorings with a single telescope, trying to estimate 
delays between the image A and the other quasar images, $\Delta t_{\rm{AX}} = t_{\rm{A}} - 
t_{\rm{X}}$ (X = B, C, D), and find microlensing signals 
\citep{MScKD,2019ApJ...887..126G}\footnote{\citet{2019ApJ...887..126G} used the notation 
$\Delta t_{\rm{AX}} = t_{\rm{X}} - t_{\rm{A}}$ rather than that defined in this paper and 
\citet{MScKD}}. Here, we use the new light curves in Section \ref{sec:lcs} along with 
state-of-the-art curve-shifting algorithms to try to robustly measure the three independent 
time delays $\Delta t_{\rm{AB}}$, $\Delta t_{\rm{AC}}$, and $\Delta t_{\rm{AD}}$. At the end 
of this section, we also discuss the intrinsic and extrinsic (microlensing) variability 
of the quasar.  

There are several cross-correlation techniques to measure time delays between light curves 
containing microlensing variations \citep[e.g.][and 
references therein]{2015ApJ...800...11L}, and thus we considered two very different methods 
and models to obtain reliable results. First, we performed the AB, AC, and AD comparisons 
using the $D^2_{4,2}$ dispersion method \citep{1996A&A...305...97P}. This technique 
evaluates the dispersion between two different light curves for a range of time lags and 
flux-ratio model parameters. For given values of the time lag and model parameters, each 
data point in one of the two light curves is compared with magnitudes in the other curve at 
time separations shorter than a decorrelation length $\delta$, and squared differences 
between pairs of data with longer separations and larger photometric errors have smaller 
weights. The key idea of the method is to find the time lag and model parameters that 
minimises the dispersion, i.e. the weighted sum of squared differences.

Our flux ratio model accounted for microlensing variability by incorporating 
four magnitude offsets instead a single one. In order to match the light curve of the 
reference image A and the shifted curve of another quasar image, the curve of A was splitted 
into four one-year segments, covering one observing season each. We then assumed a constant 
magnitude offset within every segment, while the offset was allowed to vary from one segment 
to other. This seasonal magnitude offset (SMO) model works well in presence of intra-year 
microlensing events and microlensing variability on timescales of several years 
\citep{2016A&A...596A..77G,2021A&A...646A.165S}. We have written a Python code to minimise 
the $D^2_{4,2}$(AB), $D^2_{4,2}$(AC), and $D^2_{4,2}$(AD) dispersions, and used three 
reasonable $\delta$ values of 10, 15 and 20 d. This 10-d interval of $\delta$ permits us to 
account for the intrinsic variance of the technique. Figure~\ref{fig:f2} displays the three 
dispersion spectra for the AD comparison. 

We also carried out 1000 "repetitions of the experiment" by generating 1000 synthetic light 
curves of each image, peforming AB, AC, and AD comparisons from these simulated curves, and 
obtaining the distributions of time lags and magnitude offsets that minimise dispersions 
\citep[e.g.][and references therein]{2019ApJ...887..126G}. For example, the 1$\sigma$ 
confidence intervals for $\Delta t_{\rm{AD}}$ through the corresponding time lag 
distributions are: $-$166.1 $\pm$ 9.5, $-$166.3 $\pm$ 8.5, and $-$166.3 $\pm$ 8.4 d for 
$\delta$ = 10, 15, and 20 d, respectively. These measurements indicate that the intrinsic 
variance is well below the uncertainties, so hereafter we show only results for $\delta$ = 
15 d. The 1$\sigma$ confidence intervals for $\Delta t_{\rm{AB}}$, $\Delta t_{\rm{AC}}$, and 
$\Delta t_{\rm{AD}}$ are listed in Table~\ref{tab:t3}.

\setcounter{table}{2}
\begin{table}[h!]
\begin{center}
\caption{Time delays of PS J0147+4630.}
\label{tab:t3}
\begin{tabular}{lccc}
   \hline \hline
   Method/model & $\Delta t_{\rm{AB}}$ & $\Delta t_{\rm{AC}}$ & $\Delta t_{\rm{AD}}$\\
   \hline 
   $D^2_{4,2}$/SMO & $-$1.0 $\pm$ 1.8 & +4.5 $\pm$ 3.6 & $-$166.3 $\pm$ 8.5\\
   $\chi^2$/FKS    & $+$0.9 $\pm$ 4.2 & +5.4 $\pm$ 4.4 & $-$168.7 $\pm$ 6.2\\
   Combined        &      0 $\pm$ 3 &     +5 $\pm$ 4   & $-$167.5 $\pm$ 7.4\\
   \hline
\end{tabular}
\end{center}
\footnotesize{Note: When applying the $D^2_{4,2}$ dispersion, we consider a SMO model 
without assumptions about the intrinsic variability. We also use the $\chi^2$ technique 
along with a FKS model to describe the intrinsic and extrinsic variations. Additionally, we 
combine $D^2_{4,2}$/SMO and $\chi^2$/FKS delays in a simple way, i.e. calculating mean 
central values and mean errors. We adopt these averages for subsequent studies. Here, 
$\Delta t_{\rm{AX}}$ (X = B, C, D) are in days, image A leads image X if $\Delta t_{\rm{AX}} 
<$ 0 (otherwise A trails X), and all measurements are 68\% confidence intervals.}
\end{table}  

Second, the time delays of \object{PS J0147+4630} were inferred from PyCS3 curve-shifting 
algorithms\footnote{https://gitlab.com/cosmograil/PyCS3} \citep{2013A&A...553A.120T,PyCS3,
2020A&A...640A.105M}. PyCS3 is a software toolbox to estimate time delays between images of 
gravitationally lensed quasars, and we focused on the $\chi^2$ technique, assuming that the
intrinsic signal and the extrinsic ones can be modelled as a free-knot spline (FKS). This 
technique shifts the four light curves simultaneously (ABCD comparison) to better constrain 
the intrinsic variability, and relies on an iterative nonlinear procedure to fit the four 
time shifts and splines that minimise the $\chi^2$ between the data and model 
\citep{2013A&A...553A.120T}. Results depend on the initial guesses for the time shifts, so 
it is necessary to estimate the intrinsic variance of the method using a few hundred initial 
shifts randomly distributed within reasonable time intervals. In addition, a FKS is 
characterised by a knot step, which represents the initial spacing between knots. The model 
consists of an intrinsic spline with a knot step $\eta$ and four independent extrinsic 
splines with $\eta_{\rm{ml}}$ that account for the microlensing variations in each quasar 
image \citep{2020A&A...640A.105M}. 

To address the intrinsic variability, we considered three $\eta$ values of 30, 50 and 70 d. 
Intrinsic knot steps shorter than 30 d fit the observational noise, whereas $\eta$ values 
longer than 70 d do not fit the most rapid variations of the source quasar. Intrinsic 
variations are usually faster than extrinsic ones, and additionally, the software works fine 
when the microlensing knot step is significantly longer than $\eta$. Therefore, the 
microlensing signals were modelled as free-knot splines with $\eta_{\rm{ml}}$ = 350$-$400 d 
\citep[i.e. values intermediate between those shown in Table 2 of][]{2020A&A...640A.105M}. 
We also generated 500 synthetic (mock) light curves of each quasar image, optimised every 
mock ABCD dataset, and checked the similarity between residuals from the fits to the 
observed curves and residuals from the fits to mock curves. The comparison of residuals was 
made by means of two statistics: standard deviation and normalised number of runs 
$Z_{\rm{r}}$ \citep[see details in][]{2013A&A...553A.120T}. For $\eta$ = 50 d and 
$\eta_{\rm{ml}}$ = 400 d, histograms of residuals derived from mock curves (grey) and from 
the LT-NOT light curves of \object{PS J0147+4630} are included in the top panels of 
Figure~\ref{fig:f3}. It is apparent that the standard deviations through the synthetic and 
the observed curves match very well. Additionally, the bottom panels of Figure~\ref{fig:f3} 
show distributions of $Z_{\rm{r}}$ from synthetic light curves (grey) for $\eta$ = 50 d and 
$\eta_{\rm{ml}}$ = 400 d. These bottom panels also display the $Z_{\rm{r}}$ values from the 
observations (vertical lines), which are typically located at $\sim$0.3$\sigma$ of the mean 
values of the synthetic distributions. 

Four pairs of ($\eta$, $\eta_{\rm{ml}}$) values (see above) led to the set of time delays in 
Figure~\ref{fig:f4}. We have verified that other feasible choices for $\eta_{\rm{ml}}$ (e.g. 
$\eta_{\rm{ml}}$ = 200 d) do not substantially modify the results in this figure. The black 
horizontal bars correspond to 1$\sigma$ confidence intervals after a marginalisation over 
results for all pairs of knot steps, and those in the left panels of Figure~\ref{fig:f4} are 
included in Table~\ref{tab:t3}. We finally adopted the time delays in the fourth row of 
Table~\ref{tab:t3}, which were obtained by averaging central values and errors in the two 
previous rows. It seems to be difficult to accurately determine delays between the brightest 
images ABC because they are really short. To robustly measure $\Delta t_{\rm{AB}}$ and 
$\Delta t_{\rm{AC}}$ in a near future, we will most likely need to follow a non-standard 
strategy focused on several time segments associated with strong intrinsic variations and 
weak extrinsic signals. Fortunately, we find an accurate and reliable value of $\Delta 
t_{\rm{AD}}$ (uncertainty of about 4\%), confirming the early result by \citet{MScKD}. 

\begin{figure*}
\centering
\includegraphics[width=9cm]{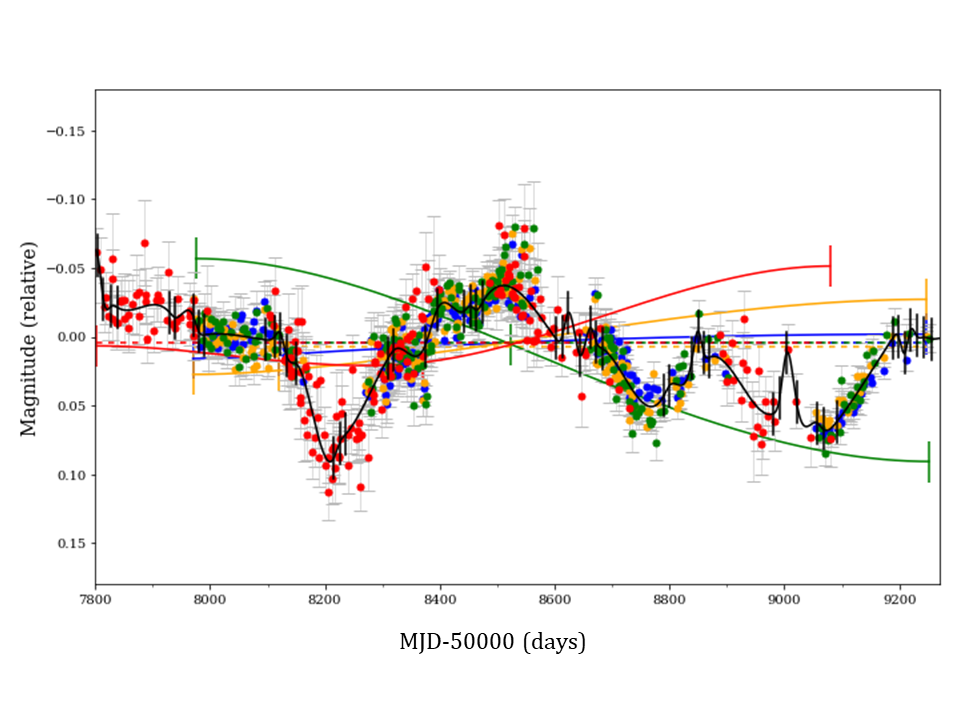}
\includegraphics[width=9cm]{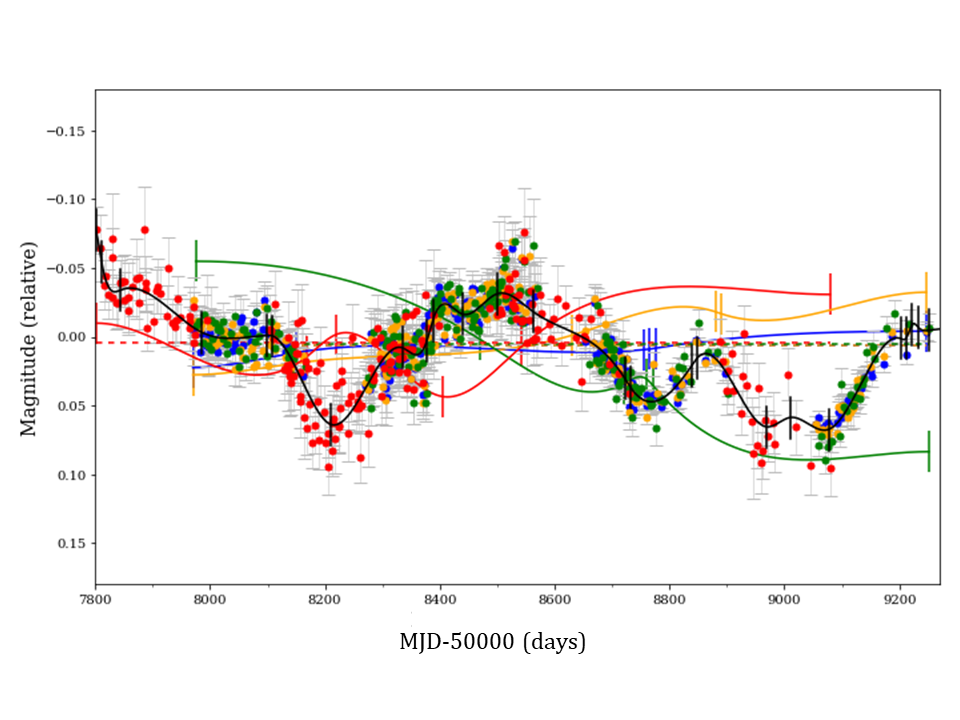}
\caption{Free-knot spline fits of the LT-NOT data of PS J0147+4630. The black lines 
depict the intrinsic splines, and the blue, orange, green, and red lines represent the 
splines for the microlensing variations of A, B, C, and D, respectively. Knots are shown as 
vertical ticks, whereas the horizontal dashed lines are references for microlensing. The 
intrinsic splines trace reasonably well the ensembles of circles: microlensing-corrected and 
median-subtracted light curves of the reference image A (blue), along with 
microlensing-corrected, time-shifted, and median-subtracted light curves of B (orange), C 
(green), and D (red). The curves of BCD are shifted in time by the adopted time delays (see 
Table~\ref{tab:t3}), and all curves are corrected from their microlensing splines. {\it 
Left}: $\eta$ = 30 d, $\eta_{\rm{ml}}$ = 400 d. {\it Right}: $\eta$ = 70 d, $\eta_{\rm{ml}}$ 
= 200 d.}
\label{fig:f5}
\end{figure*}

\begin{figure*}
\centering
\includegraphics[width=9cm]{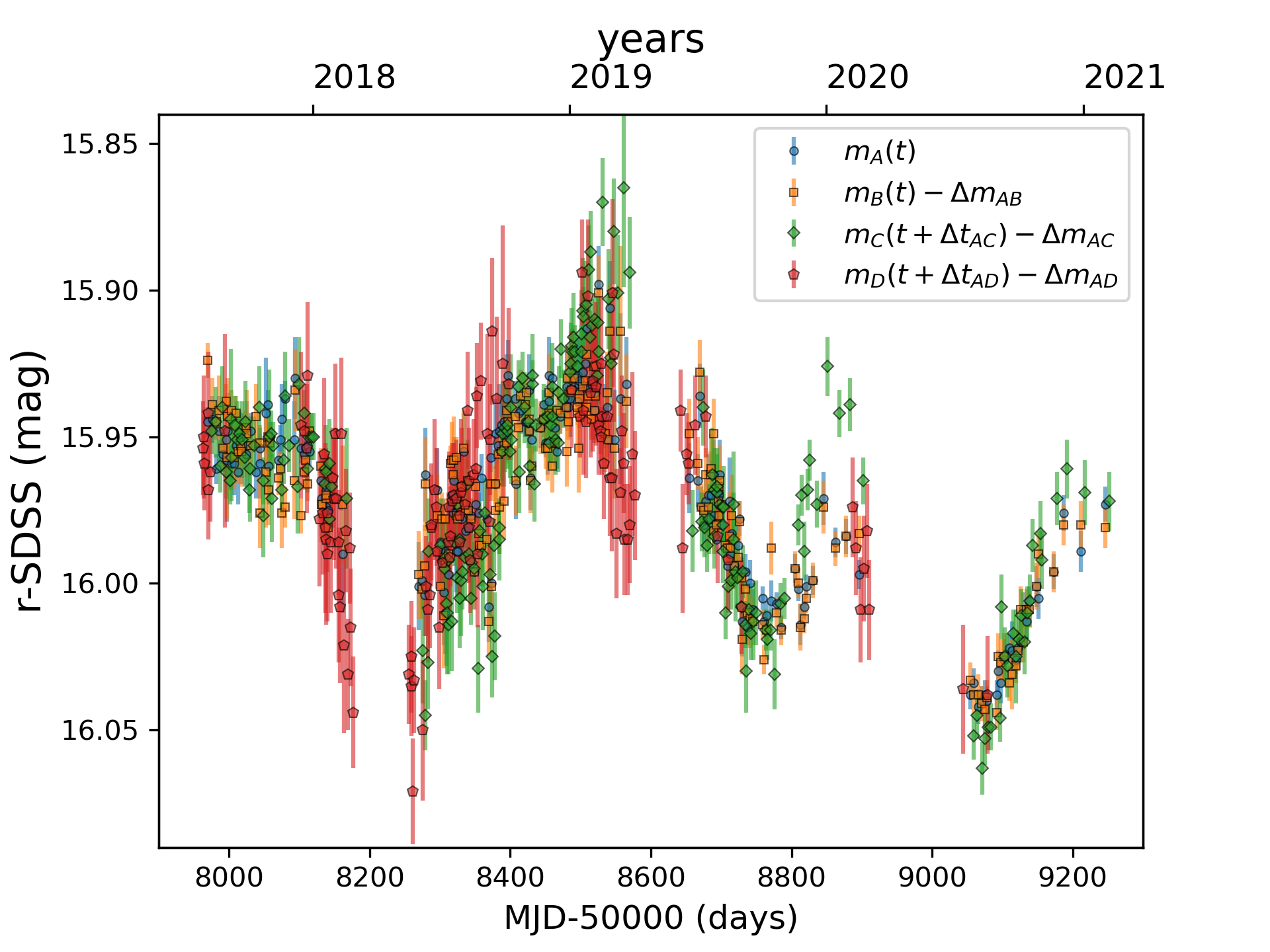}
\includegraphics[width=9cm]{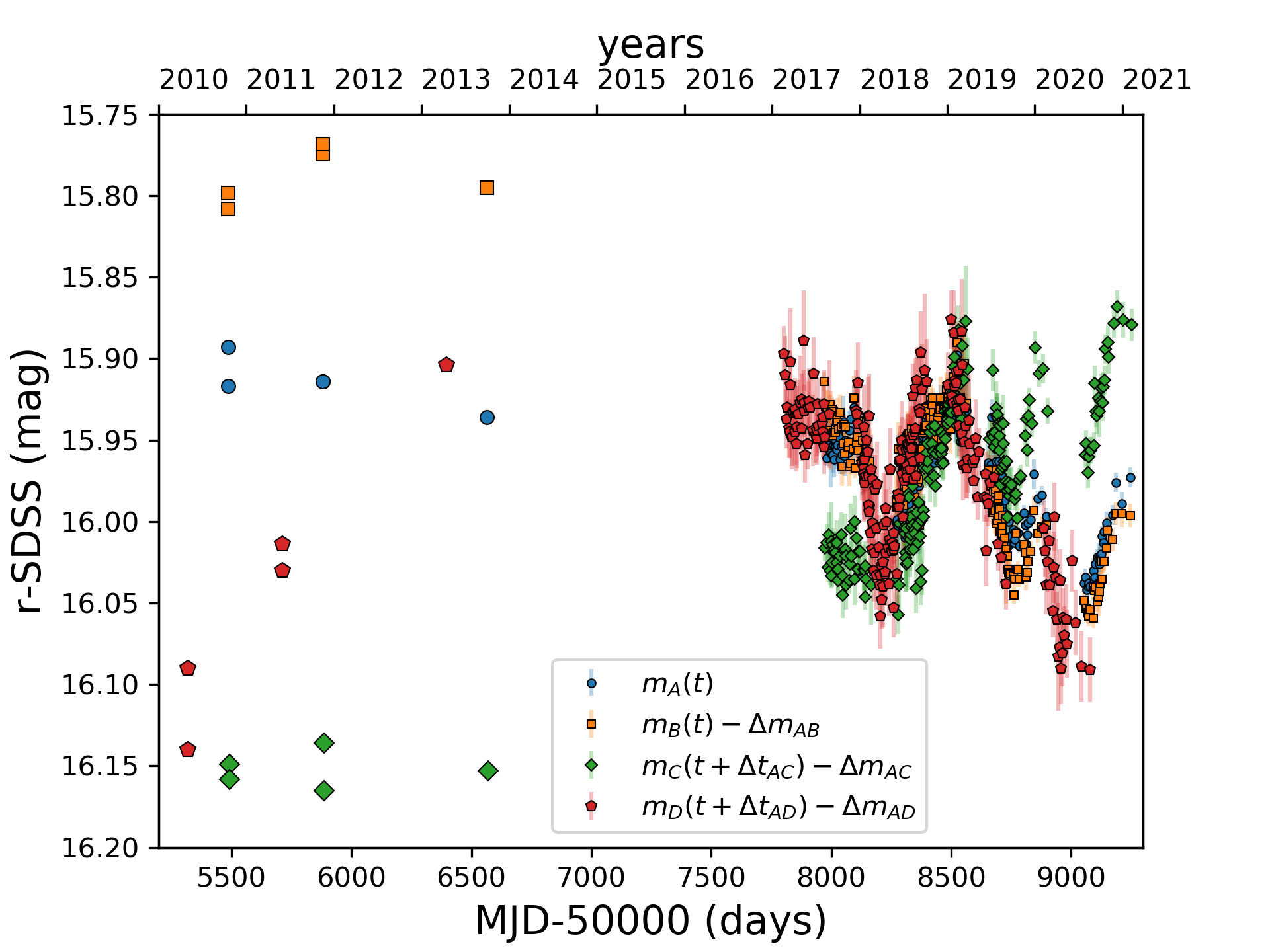}
\caption{Shifted light curves in the $r$ band. {\it Left}: LT-NOT data. Overlapping 
between A original data and shifted data of BCD. To properly shift the BCD light curves, we 
consider the adopted time delays and corresponding seasonal magnitude offsets from the SMO 
model (see main text). {\it Right}: LT-NOT data plus photometric data from Pan-STARRS 
frames in 2010$-$2013. The original brightness record of A is compared with shifted light 
curves of B ($\Delta m_{\rm{AB}}$ = $-$0.264 mag), C ($\Delta t_{\rm{AC}}$ = +5 d, $\Delta 
m_{\rm{AC}}$ = $-$0.590 mag), and D ($\Delta t_{\rm{AD}}$ = $-$167.5 d, $\Delta m_{\rm{AD}}$ 
= $-$2.304 mag). The curve of B is not shifted in time because $\Delta t_{\rm{AB}}$ = 0 (see 
main text for details).}
\label{fig:f6}
\end{figure*}

After building median-subtracted light curves, the central values of the adopted time 
delays (see Table~\ref{tab:t3}) were used to shift in time such normalised curves. As A is 
the reference image and $\Delta t_{\rm{AB}}$ = 0, the new curves of A and B retained their 
original epochs. We then applied the PyCS3 software ($\chi^2$/FKS model) to simultaneously 
fit the intrinsic signal and the microlensing variability of each quasar image. 
Figure~\ref{fig:f5} displays our results for $\eta$ = 30 d and $\eta_{\rm{ml}}$ = 400 d 
(left panel), and $\eta$ = 70 d and $\eta_{\rm{ml}}$ = 200 d (right panel). The black lines 
with knot vertical ticks model the intrinsic variation shared by the four light curves. Both 
intrinsic signal reconstructions are compared with microlensing-corrected and time-shifted 
normalised curves (circles). The blue, orange, green, and red lines model the microlensing 
variations of A, B, C, and D, respectively. The A image is weakly affected by microlensing, 
while the other three images show microlensing episodes with total amplitudes exceeding 0.05 
mag, and the extrinsic variation of C is particularly prominent. We note that the two pairs 
of ($\eta$, $\eta_{\rm{ml}}$) values we use in Figure~\ref{fig:f5} lead to similar delays 
between A and D (both consistent with the adopted one), but they produce microlensing 
splines for D (red lines) having different behaviours on time scales of hundreds of days.

\setcounter{table}{4}
\begin{table*}
\begin{center}
\caption{Redshift and single-epoch image flux ratios of PS J0147+4630.}
\label{tab:t5}
\begin{tabular}{lcccc}
   \hline \hline
   Emission & $z_{\rm{s}}$ & $B/A$ & $C/A$ & $D/A$\\
   \hline 
   Mg\,{\sc ii} 			&  2.355 &  $-$  &  $-$  & 0.065 $\pm$ 0.001\\
   cont@2800    			&   $-$  &  $-$  &  $-$  & 0.113 $\pm$ 0.001\\
                                        &        &       &       &      \\
   H$\beta$				&   $-$  & 0.450 $\pm$ 0.012 & 0.263 $\pm$ 0.010 & 0.065 $\pm$ 0.003\\
   \big[O\,{\sc iii}\big] 		&  2.357 &  $-$  &  $-$  &  $-$ \\
   \big[O\,{\sc iii}\big]$\lambda 5007$ &   $-$  & 0.559 $\pm$ 0.038 & 0.311 $\pm$ 0.037 & 0.072 $\pm$ 0.017\\
   cont@5100 				&   $-$  & 0.593 $\pm$ 0.012 & 0.423 $\pm$ 0.008 & 0.122 $\pm$ 0.005\\
                                     	&        &       &       &      \\
   H$\alpha$ main			&  2.359 & 0.484 $\pm$ 0.003 & 0.292 $\pm$ 0.002 & 0.064 $\pm$ 0.001\\
   H$\alpha$ VBC	 		&   $-$  & 0.541 $\pm$ 0.007 & 0.406 $\pm$ 0.005 & 0.068 $\pm$ 0.003\\
   cont@6563 				&   $-$  & 0.636 $\pm$ 0.002 & 0.435 $\pm$ 0.001 & 0.152 $\pm$ 0.001\\
   \hline
\end{tabular}
\end{center}
\footnotesize{Note: The Mg\,{\sc ii} and cont@2800 (continuum at $\lambda_{\rm{rest}}$ = 
2800 \AA) emissions are derived from the multi-component decomposition of Keck-ESI spectra 
(see the right panel of Figure~\ref{fig:f7}). The H$\beta$, [O\,{\sc iii}], cont@5100 
(continuum at $\lambda_{\rm{rest}}$ = 5100 \AA), H$\alpha$, and cont@6563 (continuum at 
$\lambda_{\rm{rest}}$ = 6563 \AA) emissions are inferred from multi-component decompositions 
of GTC-EMIR spectra (see Figures~\ref{fig:f10} and \ref{fig:f11}).}
\end{table*}

We also considered the adopted time delays to fit seasonal magnitude offsets between 
image A and the other three images ($D^2_{4,2}$/SMO model). In the left panel of 
Figure~\ref{fig:f6}, we show the overlaps between the A data and the magnitude- and 
time-shifted curves of BCD. It is noteworthy that there is a significant overlap between the 
original curve of A and the shifted curve of D (see also Figure~\ref{fig:f5}), supporting 
the reliability of the $\Delta t_{\rm{AD}}$ measurements in Table~\ref{tab:t3}. In addition 
to an image comparison spanning four years, a comparison over an 11-year period is depicted 
in the right panel of Figure~\ref{fig:f6}. We have downloaded five $r$-band warp frames of 
\object{PS J0147+4630} that are included in the Data Release 2 of the Pan-STARRS. These 
Pan-STARRS frames were obtained on three nights in the 2010$-$2013 period, i.e. a few years 
before the discovery of the lens system. Two frames are available on two of the three 
nights, so rough photometric uncertainties through average intranight variations are 0.012 
(A), 0.008 (B), 0.019 (C), and 0.033 (D) mag. To discuss the differential microlensing 
variability of the images BCD with respect to A, the right panel of Figure~\ref{fig:f6} 
shows the original curve of A along with shifted curves of BCD. We used the adopted time 
delays and arbitrary (constant) magnitude offsets to shift curves. The shapes of the four 
brightness records indicate the presence of long-term microlensing effects and suggest that 
\object{PS J0147+4630} is a suitable target for a deeper analysis of its microlensing 
signals.

\section{Quasar redshift and image flux ratios from recent spectroscopic data}
\label{sec:spec}

\citet{2018ApJ...859..146R} estimated the redshift of \object{PS J0147+4630} by 
cross-correlating a quasar spectral template with spectra of the four quasar images at 
wavelengths shorter than 5500 \AA. This blue spectral region contains emission lines that 
are severely affected by absorption features, and even excluding the Ly$\alpha$ and N\,{\sc 
v} lines, Rubin et al. obtained an inaccurate redshift $z_{\rm{s}}$ = 2.377 $\pm$ 0.007 for 
the BAL quasar. The C\,{\sc iii}] emission of the broad absorption-line quasar is observed 
at $\sim$6400 \AA, and \citet{2017A&A...605L...8L} measured $z_{\rm{s}}$ = 2.341 $\pm$ 0.001 
using only such emission line, which is apparently free of significant absorption-induced 
distortions. In this section, we analyse several emission lines at longer wavelengths, 
checking the reliability of the current value of $z_{\rm{s}}$ and trying to get information 
on image flux ratios.

The Keck Observatory Archive includes relevant data of \object{PS J0147+4630} on 1 December 
2018 (MJD$-$50\,000 = 
8453)\footnote{\url{https://koa.ipac.caltech.edu/cgi-bin/KOA/nph-KOAlogin} (Program ID: 
U122, Program PI: C. Fassnacht)}. These deep spectroscopic observations with the Echellette 
Spectrograph and Imager \citep[ESI;][]{2002PASP..114..851S} consisted of 3$\times$2400 s 
exposures using an 1\farcs0-width slit with a spatial pixel scale of 0\farcs154. The slit 
was oriented along the line joining A and D. In addition, the ESI wavelength range and its 
resolving power were 3900$-$10\,500 \AA\ and 4000, respectively. We downloaded the exposures 
of the lens system and spectroscopic data of the standard star Feige110, as well as CuAr, 
Xe, and HgNe lamps exposures for wavelength calibration. Data reduction and spectral 
extraction were then performed using the MAuna Kea Echelle Extraction (MAKEE) package by Tim 
Barlow\footnote{\url{https://sites.astro.caltech.edu/~tb/makee}}. To extract the individual 
spectra of A and D from the MAKEE software, we considered two apertures with 7 pixel size,
which are similar to the slit width and the FWHM seeing. The normalised spectrum of the 
standard star allowed us to calibrate in flux and correct by telluric absorption the quasar 
spectra. 

\begin{figure*}
\centering
\includegraphics[width=9cm]{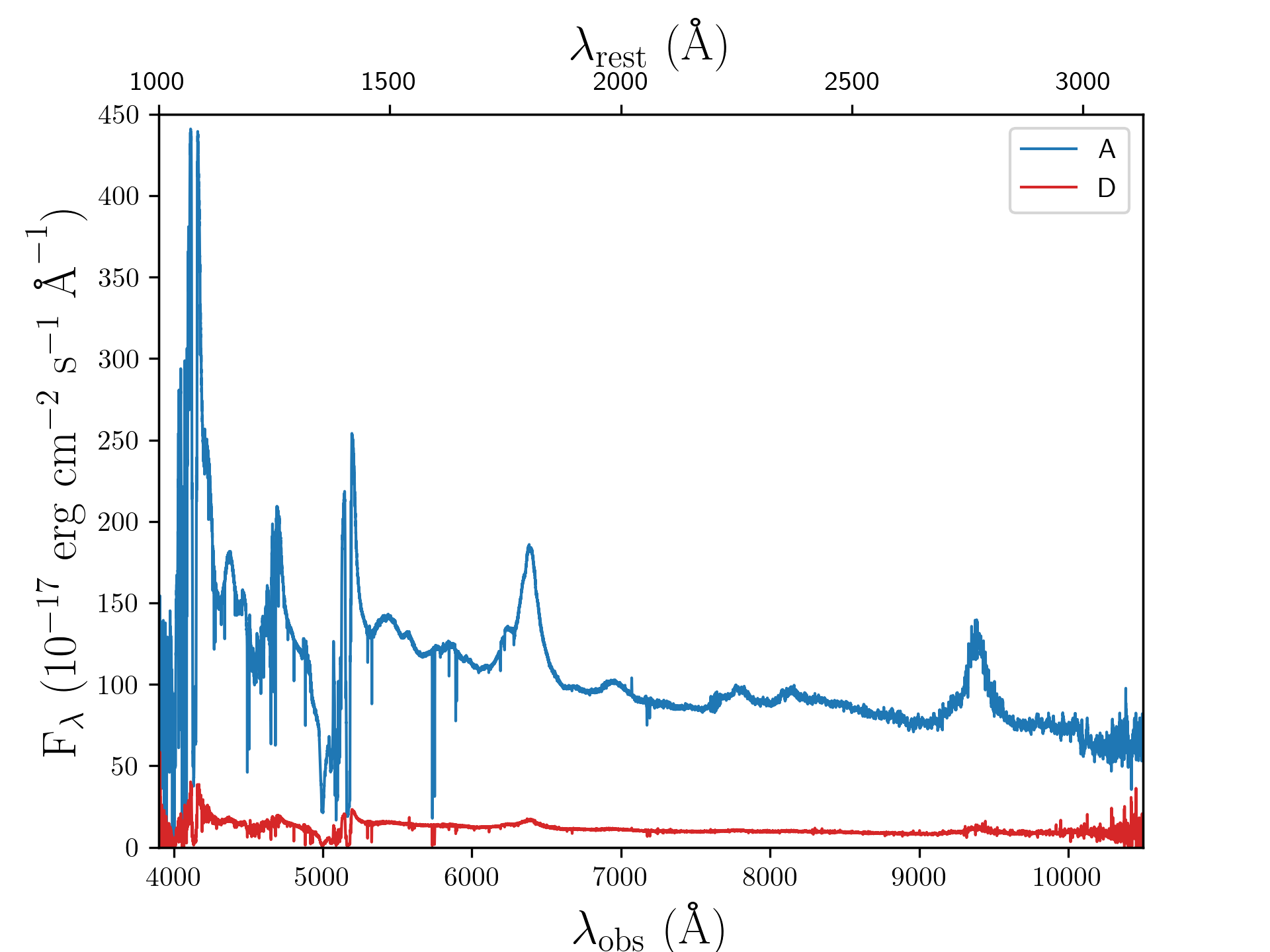}
\includegraphics[width=9cm]{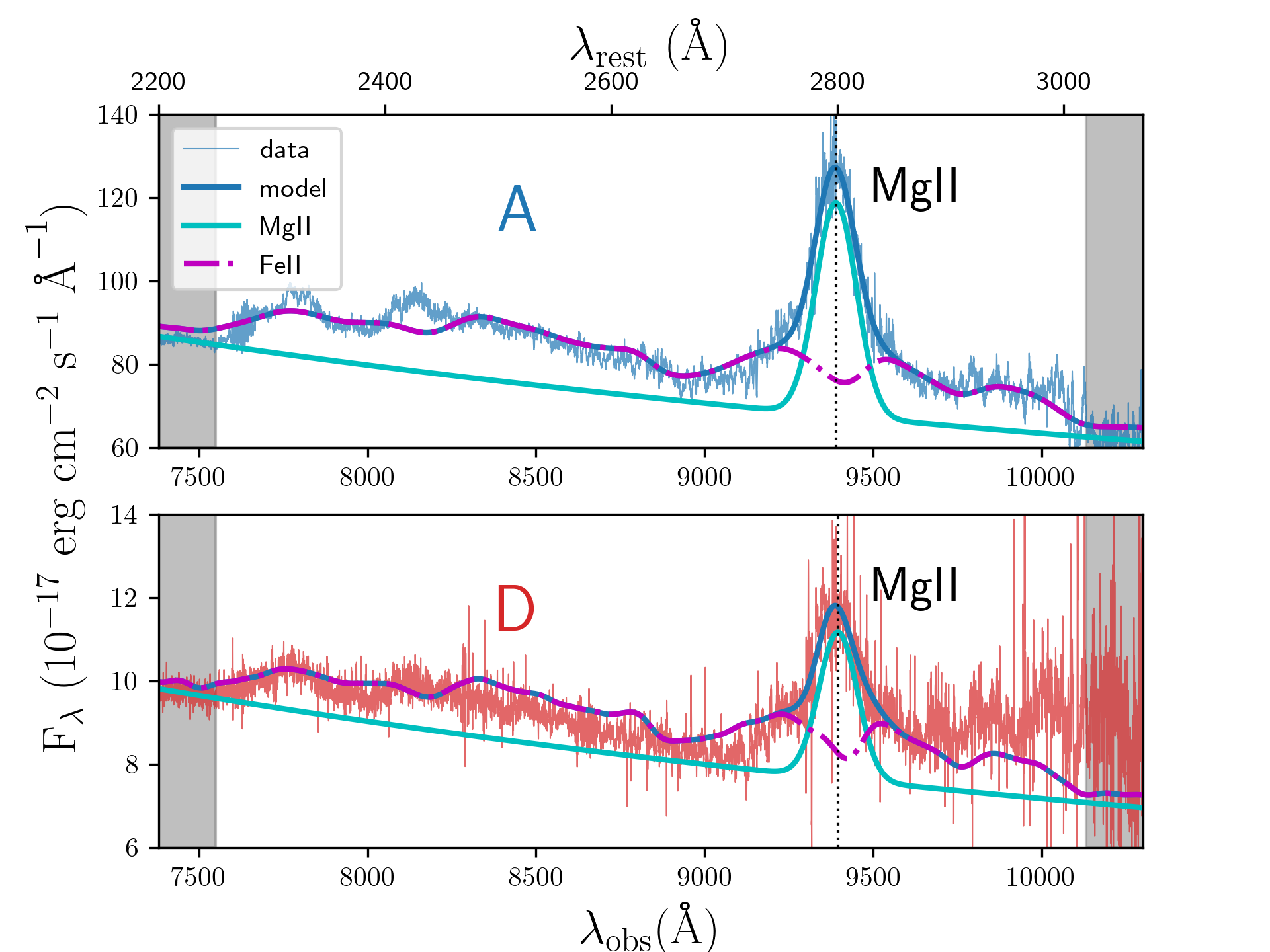}
\caption{{\it Left}: Keck-ESI spectra of PS J0147+4630AD in December 2018. {\it Right}:  
multi-component decomposition of the Mg\,{\sc ii} line spectral region in the Keck-ESI 
spectra (see main text). The pure Mg\,{\sc ii} emission is modelled as a Gaussian function, 
and the two Gaussian distributions (A and D) with rest-frame central wavelengths of 
$\sim$2800 \AA\ (vertical dotted lines) are drawn superimposed on power-law continua.}
\label{fig:f7}
\end{figure*}

The final spectra of A and D are available in tabular format at the CDS: Table 4 includes 
fluxes of both quasar images covering the spectral range 3900$-$10\,500 \AA\ with 33\,001 
channels of 0.2 \AA\ each. The Keck-ESI spectra of the two images are also plotted in the 
left panel of Figure~\ref{fig:f7}. These new spectra show emission lines that were 
previously detected at visible wavelengths \citep{2017A&A...605L...8L,2018ApJ...859..146R} 
and the Mg\,{\sc ii} line at $\sim$9400 \AA. Thus, we focused on the analysis of the 
Mg\,{\sc ii} line spectral region using a decomposition into three components: power-law 
continuum, Fe\,{\sc ii} pseudo-continuum, and Gaussian Mg\,{\sc ii} emission (see the right 
panel of Figure~\ref{fig:f7}). The two Gaussian distributions (A and D) led to an unbiased 
redshift $z_{\rm{s}}$ = 2.355. Additionally, the $D/A$ flux ratios for the pure Mg\,{\sc ii} 
emission and the continuum at $\lambda_{\rm{rest}}$ = 2800 \AA\ are given in 
Table~\ref{tab:t5}. Uncertainties in these flux ratios (1$\sigma$ confidence intervals)
were estimated from 300 repetitions of each spectrum. To obtain synthetic spectra for A and 
D, we modified the observed fluxes by adding realizations of normal distributions around 
zero, with standard deviations equal to the measured errors. 

\begin{figure}
\centering
\includegraphics[width=9cm]{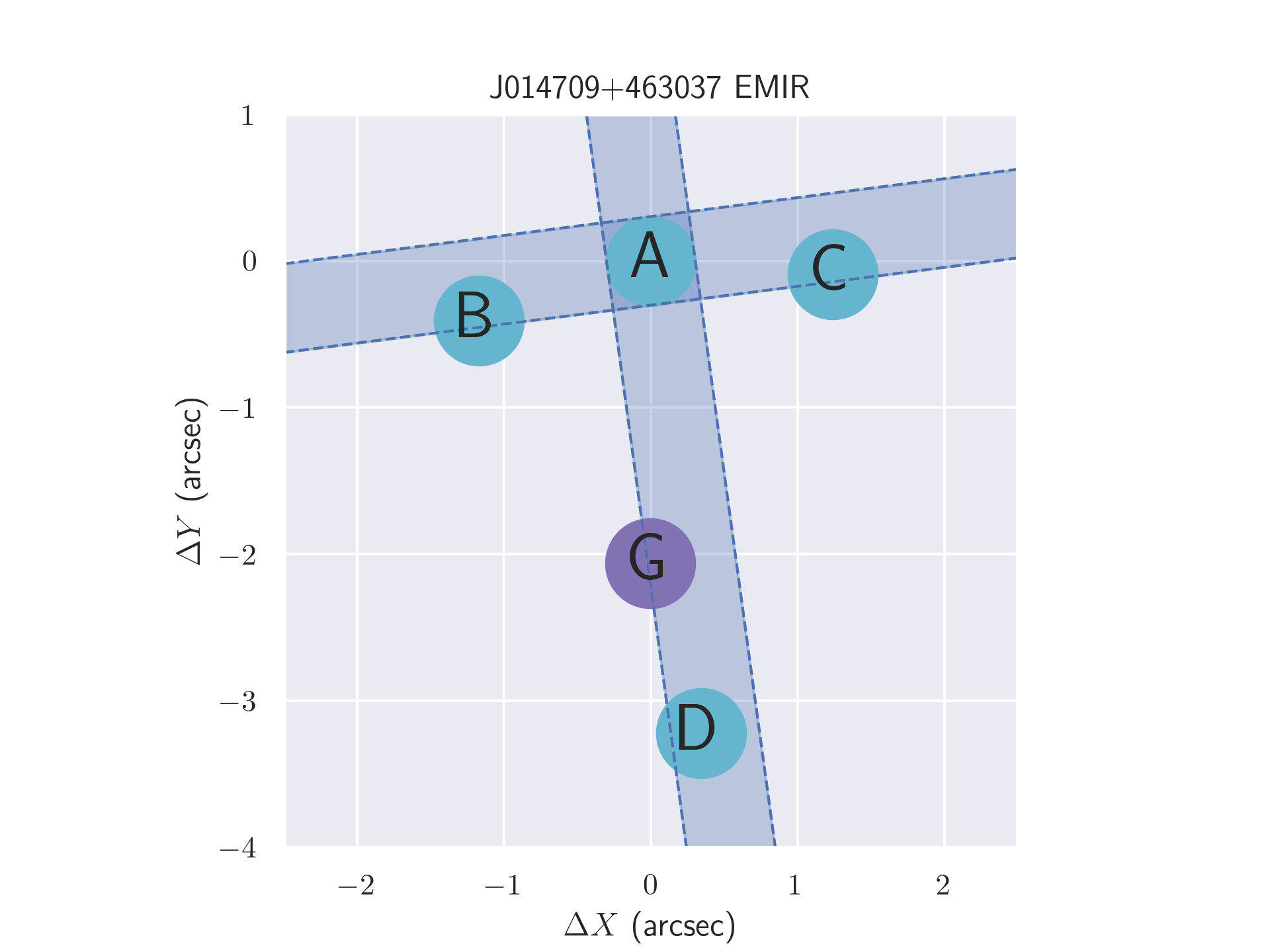}
\caption{Positions of the four images and the lens galaxy of PS J0147+4630 on the two slit 
orientations for GTC-EMIR. Using the almost horizontal slit (orientation 1), it is possible 
to obtain spectra of ABC. The almost vertical slit (orientation 2) is suitable for 
extracting spectra of ADG. The circles represent seeing discs (diameter = FWHM seeing).}
\label{fig:f8}
\end{figure}

Recent spectroscopic observations of \object{PS J0147+4630} with the near-IR instrument EMIR 
\citep{2016SPIE.9908E..1JG} on the 10.4 m Gran Telescopio Canarias (GTC) are useful tools to
discuss the H$\beta$, [O\,{\sc iii}], and H$\alpha$ emissions from the distant quasar. These
data were taken on 14 August 2019 (MJD$-$50\,000 = 8709) under excellent seeing conditions 
(FWHM seeing $\sim$ 0\farcs6) and are available at the GTC Public 
Archive\footnote{\url{https://gtc.sdc.cab.inta-csic.es/gtc/} (Program ID: GTCMULTIPLE2B-19B, 
Program PI: R. Scarpa)}. The 0\farcs6-width slit was oriented along two different 
directions (see Figure~\ref{fig:f8}): one crossing the ABC images (slit orientation 1; total 
exposure time of 1920 s = 12$\times$160 s), and the other crossing the AD images and the 
lens galaxy G (slit orientation 2; total exposure time of 3840 s = 24$\times$160 s). The 
spatial pixel scale was 0\farcs1915, while the EMIR-$HK$ pseudo-grism wavelength range and 
resolving power were 1.45$-$2.42 $\mu$m and 987, respectively. There are also observations 
of the standard telluric star HIP10814 on the same night.

\begin{figure}
\centering
\includegraphics[width=9cm]{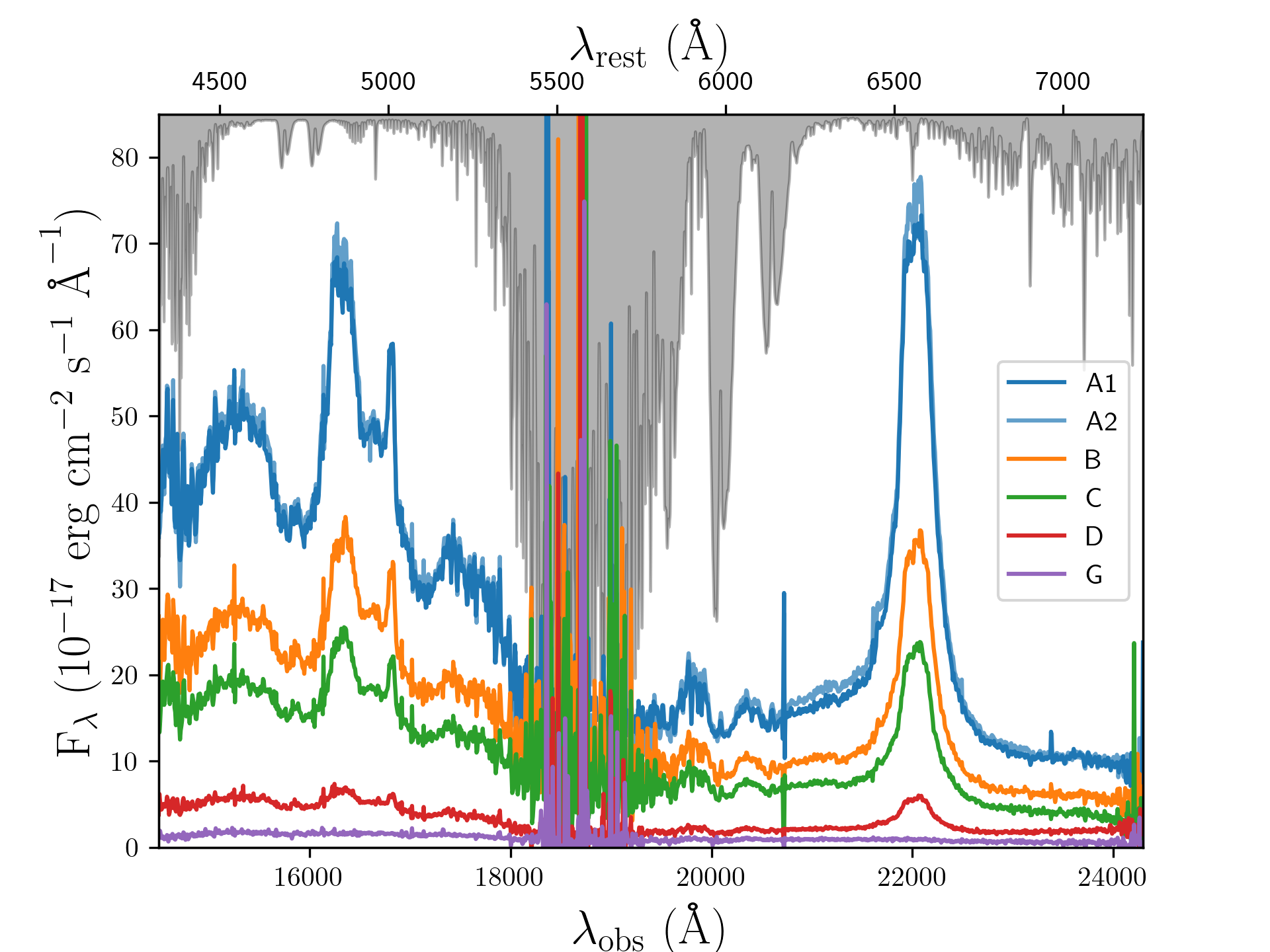}
\caption{GTC-EMIR spectra of PS J0147+4630 in August 2019. Spectra A1, B, and C correspond 
to the slit orientation 1, and spectra A2, D, and G are associated with the slit orientation 
2 (see Figure~\ref{fig:f8}). The grey line describes the atmospheric transmission. Water and 
carbon dioxide completely block transmission of light at $\sim$1.9 $\mu$m.}
\label{fig:f9}
\end{figure}

\begin{figure}
\centering
\includegraphics[width=9cm]{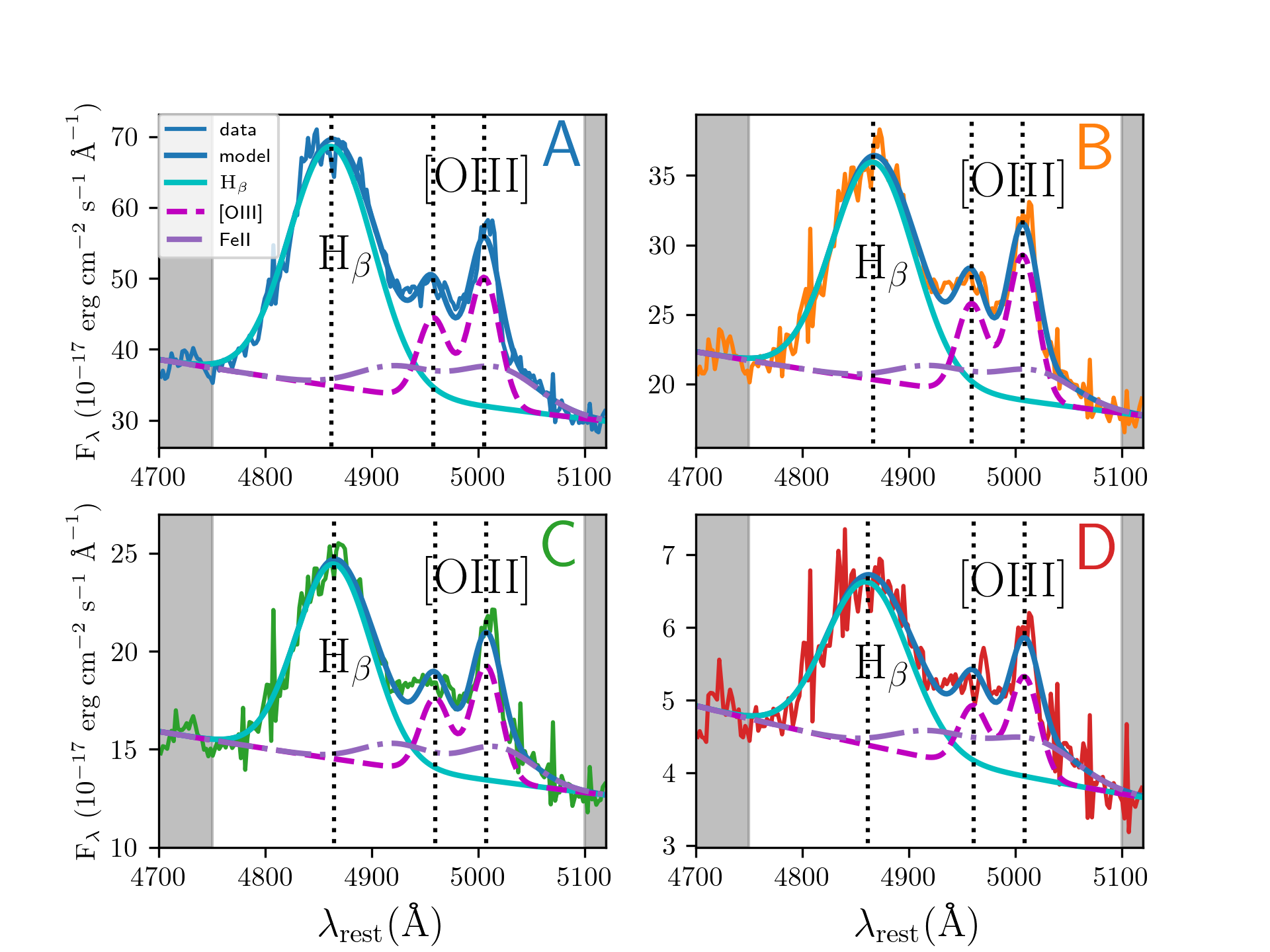}
\caption{Multi-component decomposition of the H$\beta$+[O\,{\sc iii}] complex spectral 
region in the GTC-EMIR spectra (the spectrum of A is the average of A1 and A2 in 
Figure~\ref{fig:f9}; see main text). Each of the H$\beta$, [O\,{\sc iii}]$\lambda 4959$, and 
[O\,{\sc iii}]$\lambda 5007$ emissions is modelled as a Gaussian function. All Gaussian 
distributions are drawn superimposed on power-law continua, and their rest-frame central 
wavelengths are highlighted with vertical dotted lines.}
\label{fig:f10}
\end{figure}

\begin{figure}
\centering
\includegraphics[width=9cm]{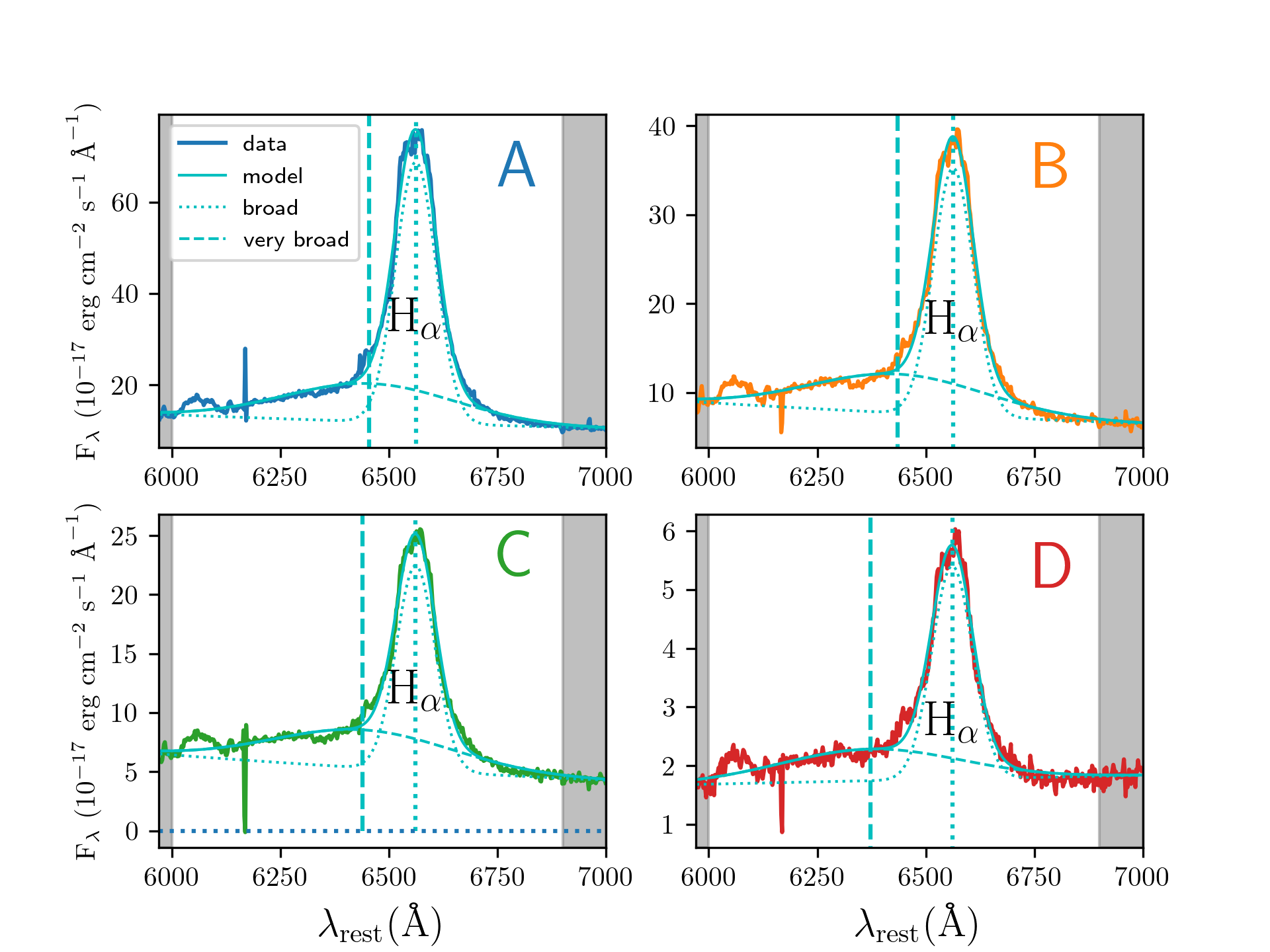}
\caption{Multi-component decomposition of the H$\alpha$ line spectral region in the GTC-EMIR 
spectra (the spectrum of A is the average of A1 and A2 in Figure~\ref{fig:f9}; see main 
text). The main (broad component) and VBC H$\alpha$ emissions are both modelled as a 
Gaussian function. The Gaussian distributions are drawn superimposed on power-law continua, 
and their rest-frame central wavelengths are highlighted with dotted vertical lines (main 
component) and dashed vertical lines (VBC).}
\label{fig:f11}
\end{figure}

We downloaded the GTC-EMIR data and used the PyEMIR software \citep{2019hsax.conf..605C} for 
their reduction. For each slit orientation, we extracted three individual spectra (ABC or 
ADG) by fitting three 1D Moffat profiles in the spatial direction for each wavelength bin 
\citep[e.g.][]{2007A&A...468..885S,2017ApJ...836...14S,2019ApJ...887..126G}. To fit the 
Moffat profiles, the positions of BCDG with respect to A were set from the HST astrometry of 
the system \citep{2019MNRAS.483.5649S,2021MNRAS.501.2833S}. Spectra were calibrated in flux, 
and telluric absorption was properly corrected. In addition, we have taken into account slit 
losses (relative to A) of 0.928 (B), 0.932 (C), and 0.995 (D) because the BCD images are not 
centred on the slit axis (see Figure~\ref{fig:f8}). An 1D Moffat profile does not account 
for the total light of the very faint galaxy G and the slit loss of G is not taken into 
account either. In this study, we are interested in the quasar spectra and warn that fluxes 
of G are underestimated. Final spectra for the slit orientations 1 and 2 are shown in Tables 
6 and 7, respectively. Both tables are available at the CDS and are structured in a similar 
way. Table 6 includes wavelengths in \AA\ and fluxes of ABC, whereas Table 7 includes the 
same wavelengths and fluxes of ADG. All fluxes are given in 10$^{-17}$ erg cm$^{-2}$ 
s$^{-1}$ \AA$^{-1}$. Figure~\ref{fig:f9} also displays the new near-IR spectra of \object{PS 
J0147+4630}. 

The GTC-EMIR quasar spectra cover the wavelength range of H$\beta$, [O\,{\sc iii}], and 
H$\alpha$ emission lines. At $\lambda_{\rm{rest}}$ around 4900 \AA, these spectra were 
modelled as the sum of a power-law continuum, a Fe\,{\sc ii} pseudo-continuum, and three 
Gaussian emissions of H$\beta$, [O\,{\sc iii}]$\lambda 4959$, and [O\,{\sc iii}]$\lambda 
5007$ (see Figure~\ref{fig:f10}). From the [O\,{\sc iii}] doublet, we determined the optimal 
redshift $z_{\rm{s}}$ = 2.357. Flux ratios were also estimated and incorporated in 
Table~\ref{tab:t5}. To account for the shape of the H$\alpha$ emission at rest-frame 
wavelengths around 6500 \AA, we considered a decomposition into three components: power-law 
continuum and two Gaussian H$\alpha$ emissions (see Figure~\ref{fig:f11}), i.e. a main 
component (the average FWHM of the velocity distributions is $\sim$4900 km s$^{-1}$) and a 
very broad component (VBC; the average FWHM velocity is $\sim$22\,300 km s$^{-1}$) that is 
blueshifted with respect to the main one by 5000$-$9000 km s$^{-1}$. The source redshift 
from the main component and several flux ratios are shown in Table~\ref{tab:t5}.
Uncertainties in flux ratios from GTC-EMIR data were derived from 300 repetitions of the 
experiment (synthetic spectra based on the observed ones; see above). To generate simulated 
spectra of a quasar image, we previously estimated spectral errors as absolute differences 
between the measured fluxes and those in a smoothed version of its observed spectrum.

Our detailed analysis of emission lines detected at near-IR wavelengths suggests that 
$z_{\rm{s}}$ = 2.357 $\pm$ 0.002 (see Table~\ref{tab:t5}). This is basically the average of 
previous redshifts based on visible spectra, and reported by \citet{2017A&A...605L...8L} and 
\citet{2018ApJ...859..146R}. The single-epoch image flux ratios in Table~\ref{tab:t5} are 
also worthy of attention. We find that all flux ratios for emission lines are smaller than 
those for continuum emissions at wavelengths inside or near wavelength ranges of lines. 
\citet{2018MNRAS.475.3086L} also indicated that the $D/A$ flux ratio for the C\,{\sc iii}] 
emission line is smaller than that for its underlying continuum. This previous result, which 
was interpreted as evidence of spectral microlensing, is fully consistent with new data in 
the last column of Table~\ref{tab:t5}. Additionally, the VBC H$\alpha$ emission is most 
likely due to an outflow in the BAL quasar, which is not detected in the H$\beta$ emission. 
We also note that possible extended blue wings of the [O\,{\sc iii}] lines cannot be 
resolved with available observations. Our data suggest a possible link between the outflow 
related to absorption features in high-ionization emission lines \citep[Ly$\alpha$, N\,{\sc 
v} and C\,{\sc iv};][]{2018ApJ...859..146R} and the VBC H$\alpha$ emission.

At the end of Section \ref{sec:mass}, flux ratio measurements in Table~\ref{tab:t5} are 
compared with macrolens flux ratios predicted by an updated lens mass model, which permits 
us to discuss extinction/microlensing scenarios. Some results of this discussion are used to 
confidently estimate the quasar black hole mass in Section \ref{sec:bhmass}.   

\section{Lens mass model}
\label{sec:mass}

\citet{2017ApJ...844...90B} presented preliminary modelling of the lens mass of \object{PS 
J0147+4630} from Pan-STARRS data, whereas \citet{2019MNRAS.483.5649S,2021MNRAS.501.2833S} 
have recently modelled the lens system using HST imaging. Shajib et al.'s solution for
the lensing mass relies on a lens scenario consisting of a singular power-law ellipsoid 
(SPLE; describing the gravitational field of the main lens galaxy G) and an external shear 
(ES; accounting for the gravitational action of other galaxies). The dimensionless surface 
mass density (convergence) profile of the SPLE was characterised by a power-law index 
$\beta$, and they found $\beta$ = 2.00 $\pm$ 0.05, where $\beta$ = 2 for an isothermal 
ellipsoid\footnote{More precisely, the original name of the power-law index in 
\citet{2019MNRAS.483.5649S,2021MNRAS.501.2833S} was $\gamma$, but we have renamed it as 
$\beta$ to avoid confusion between such index and the shear}. 

Assuming a flat $\Lambda$CDM cosmology with matter and dark energy densities of 
$\Omega_{\rm{M}}$ = 0.3 and $\Omega_{\Lambda}$ = 0.7, respectively\footnote{Results do not 
change appreciably for values of $\Omega_{\rm{M}}$ and $\Omega_{\Lambda}$ slightly different 
from those adopted here}, we first considered Shajib et al.'s solution, updated redshifts 
$z_{\rm{l}}$ = 0.678 \citep{2019ApJ...887..126G} and $z_{\rm{s}}$ = 2.357 (see Section 
\ref{sec:spec}), and our longest (most accurate) time delay in the fourth row of 
Table~\ref{tab:t3} to calculate $H_0$ and put it into perspective 
\citep[e.g.][]{2015LRR....18....2J}. Using the measured delay between images A and D, the
Hubble constant is 102 $\pm$ 11 km s$^{-1}$ Mpc$^{-1}$, in clear disagreement with currently 
accepted values around $H_0$ = 70 km s$^{-1}$ Mpc$^{-1}$. If additional mass along the line 
of sight is modelled as an external convergence $\kappa_{\rm ext}$, the factor $1 - 
\kappa_{\rm ext}$ should be $\sim$0.7 ($\kappa_{\rm ext} \sim$ 0.3) to decrease $H_0$ until 
accepted values. Therefore, the external convergence required to solve the delay crisis is 
an order of magnitude higher than typical values of $\kappa_{\rm ext}$ 
\citep[e.g.][]{2017MNRAS.467.4220R,2020A&A...643A.165B}. 

To better understand the reasons for the crisis of Shajib et al.'s solution when 
measured time delays are taken into account, the system was also modelled using astrometric 
and time-delay constraints, an SPLE + ES mass model, updated redshifts, a flat $\Lambda$CDM 
cosmology, and $H_0$ = 70 km s$^{-1}$ Mpc$^{-1}$. Thus, even in presence of a standard 
(weak) external convergence, the $H_0$ value would be consistent with accepted ones. We 
have downloaded the four HST-WFC3 frames that were used by Shajib et 
al.\footnote{\url{https://archive.stsci.edu/missions-and-data/hst} (Program ID: 15320, 
Program PI: T. Treu)}, and then determined positions of ABCDG in each of them through the 
IRAF/FITPSF/ELGAUSS task. We also considered the Gaia EDR3 positions of the quasar 
images\footnote{\url{https://gea.esac.esa.int/archive/}} \citep{2021A&A...649A...2L}. The 
Gaia-HST imaging allowed us to obtain the astrometric constraints in Table~\ref{tab:t8}, 
consisting of positions of BCDG with respect to A at the origin of coordinates. 
Table~\ref{tab:t8} shows means and standard errors of means from the five (four) independent 
solutions for the coordinates of BCD (G), so typical statistical uncertainties are 
0\farcs0008 (coordinates of BCD) and 0\farcs006 (coordinates of G).  

\setcounter{table}{7}
\begin{table}[h!]
\begin{center}
\caption{Astrometric solution from Gaia-HST data of PS J0147+4630.}
\label{tab:t8}
\begin{tabular}{lcc}
   \hline \hline
   Component & $x$ & $y$\\
   \hline 
   B	& $-$1.1683 $\pm$ 0.0003 & $-$0.4137 $\pm$ 0.0008\\
   C	& 1.2432 $\pm$ 0.0011    & $-$0.0988 $\pm$ 0.0008\\
   D	& 0.3383 $\pm$ 0.0010    & $-$3.2343 $\pm$ 0.0005\\
   G 	& 0.1632 $\pm$ 0.0055    & $-$2.0562 $\pm$ 0.0062\\	
   \hline
\end{tabular}
\end{center}
\footnotesize{Note: The A image is at the origin of coordinates (0, 0), and positive 
directions of $x$ and $y$ are defined by west and north, respectively. Both $x$ and $y$ are 
given in arcseconds.}
\end{table} 

\begin{table}[h!]
\begin{center}
\caption{Solutions for the SPLE + ES mass model of PS J0147+4630.}
\label{tab:t9}
\begin{tabular}{lcc}
   \hline \hline
   Parameter & SOC 1 & SOC 2\\
   \hline 
   $\chi^2$/d.o.f. 		& 4.0/3  	& 3.2/3             \\
   $\beta$		        & 1.83  	& 1.845 $\pm$ 0.070 \\
   $b$ (\arcsec)    		& 1.88  	& 1.882 $\pm$ 0.016 \\
   $e$ 		 		& 0.165  	& 0.151 $\pm$ 0.041 \\
   $\theta_e$ (\degr) 		& $-$71.5	& $-$67.5 $\pm$ 4.8 \\
   $\gamma$ 			& 0.174		& 0.166 $\pm$ 0.019 \\
   $\theta_{\gamma}$ (\degr)	& 10.9		& 11.5 $\pm$ 0.7    \\	
   \hline
\end{tabular}
\end{center}
\footnotesize{Note: When fitting the SPLE + ES mass model, we consider updated lens and 
source redshifts, a flat $\Lambda$CDM cosmology, $H_0$ = 70 km s$^{-1}$ Mpc$^{-1}$, and two 
different sets of observational constraints (SOC 1 and SOC 2; see main text). Position 
angles ($\theta_e$ and $\theta_{\gamma}$) are measured east of north, and $\chi^2$/d.o.f., 
$\beta$, $b$, $e$, and $\gamma$ denote reduced chi-square, power-law index, mass scale and 
ellipticity of the SPLE, and external shear strength, respectively.}
\end{table}

\begin{figure}
\centering
\includegraphics[width=9cm]{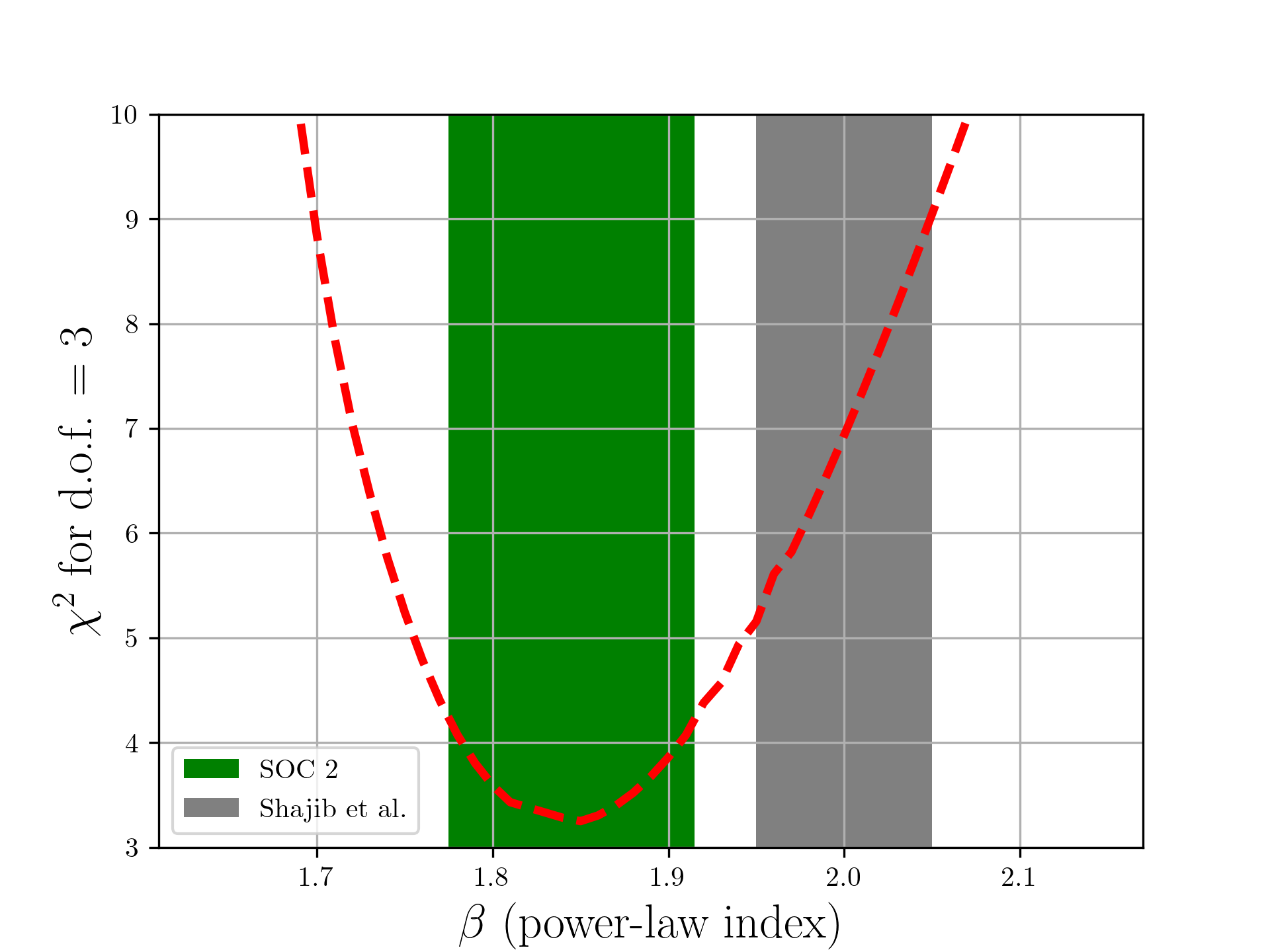}
\includegraphics[width=9cm]{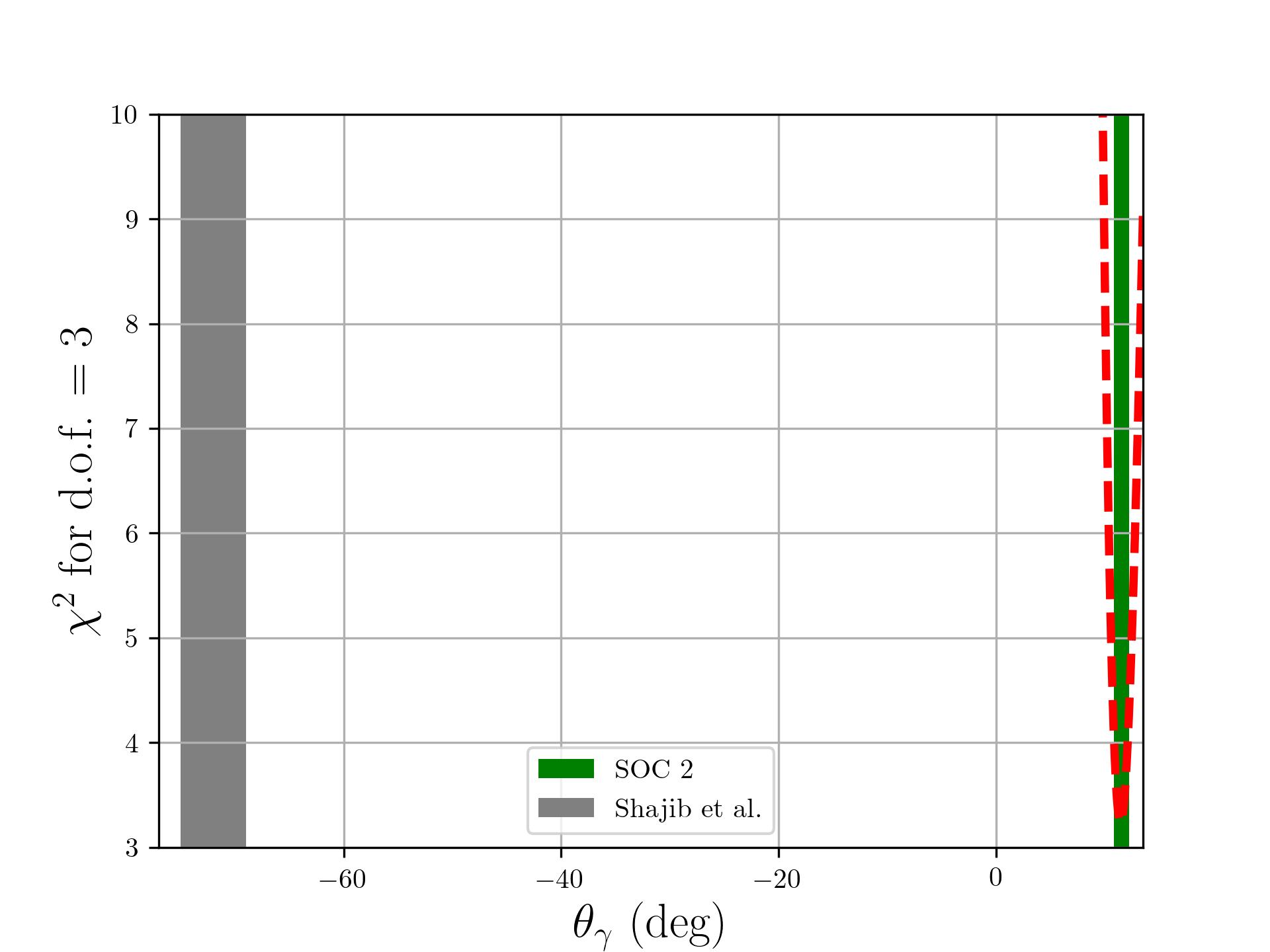}
\caption{Some results of lens mass modelling of PS J0147+4630. Our 1$\sigma$ intervals 
from the SOC 2 (see Table~\ref{tab:t9}; green rectangles) are compared with those of Shajib 
et al. (grey rectangles). {\it Top}: power-law index of the SPLE ($\beta$ = 2 for an 
isothermal distribution). {\it Bottom}: position angle of the ES.}
\label{fig:f12}
\end{figure} 

Each set of observational constraints (SOC) included the three combined time delays in 
Table~\ref{tab:t3}. In addition to these three constraints, the first set (SOC 1) 
incorporated the HST relative positions of ABCD \citep[with respect to G at the origin of 
coordinates;][]{2019MNRAS.483.5649S,2021MNRAS.501.2833S}, while the second (SOC 2) contained 
the Gaia-HST astrometry in Table~\ref{tab:t8}. When building the SOC 1, we considered the 
Shajib et al.'s astrometric uncertainties for ABCD, i.e. the original ones. In the SOC 2, we 
assumed $\sigma_x$ = $\sigma_y$ = 0\farcs0008 for BCD (see above). SPLE + ES mass models of
quads usually indicate the existence of an offset between the centre of the SPLE and the 
light centroid of the galaxy \citep[e.g.][]{2012A&A...538A..99S,2019MNRAS.483.5649S,
2021MNRAS.501.2833S}. Hence, instead of formal astrometric errors for G, we initially 
adopted $\sigma_x$ = $\sigma_y$ = 0\farcs04 in both SOC 1 and SOC 2. This uncertainty level 
equals the root-mean-square of mass/light positional offsets for most quads in the sample of 
Shajib et al. The number of observational constraints and the number of model parameters 
were 13 and 10, respectively. For three degrees of freedom (d.o.f.), the GRAVLENS/LENSMODEL 
software\footnote{\url{http://www.physics.rutgers.edu/~keeton/gravlens/}} 
\citep{2001astro.ph..2340K,gravlensmanual} led to the best fit (SOC 1) and 1$\sigma$ 
intervals (SOC 2) in Table~\ref{tab:t9}. Additionally, Figure~\ref{fig:f12} depicts the 
$\chi^2$$-$$\beta$ and $\chi^2$$-$$\theta_{\gamma}$ relationships from the SOC 2.

The SOC 1 and SOC 2 produce similar parameter values, and hereafter we focus on results 
from the SOC 2 because it leads to a better fit in terms of the reduced chi-square. 
Regarding the mass of the early-type galaxy G, it is clear that a convergence a little 
shallower than isothermal ($\beta <$ 2) is required to reasonably fit the measured time 
delays when $H_0$ = 70 km s$^{-1}$ Mpc$^{-1}$ (see Table~\ref{tab:t9} and the top panel of 
Figure~\ref{fig:f12}). Additionally, our values of the mass scale and ellipticity ($e = 1 - 
q$, where $q$ is the axis ratio) agree with those in Table 3 of Shajib et al., and the new 
orientation of G ($\theta_e$) does not differ substantially from the previous one. 
Therefore, there is still a high misalignment angle between light and mass distributions. 
This misalignment may be true or an artefact arising from the SPLE + ES scenario that we and 
Shajib et al. assumed. Despite the new external shear orientation ($\theta_{\gamma}$) is 
almost perpendicular to that of Shajib et al. (see the bottom panel of 
Figure~\ref{fig:f12}), the external shear strength around 0.166 in Table~\ref{tab:t9} 
coincides with the previous one. Most early-type galaxies reside in overdense regions, so 
external tidal fields in their vicinity are expected to have relatively high amplitudes. 
External shear strengths for quads exceeding 0.1 are consistent with N-body simulations and 
semianalytic models of galaxy formation \citep{2003ApJ...589..688H}. Using a model 
consisting of a singular isothermal elliptical potential and external shear, 
\citet{2021ApJ...915....4L} have also shown that \object{PS J0147+4630} is a shear-dominated 
system.

\begin{figure}
\centering
\includegraphics[width=9cm]{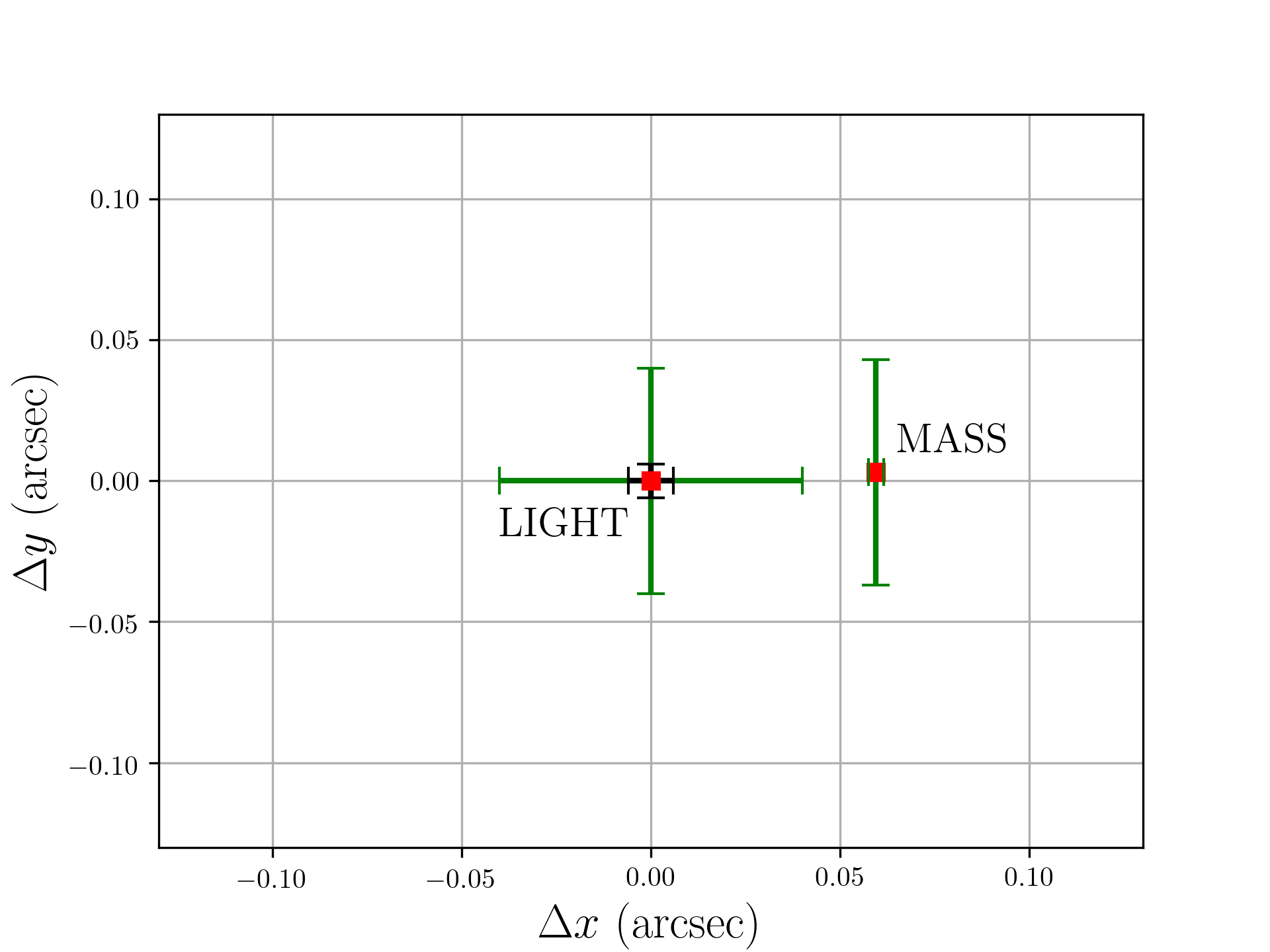}
\caption{Mass/light positional offset for the lens galaxy of PS J0147+4630. The 
observed (light) centroid of G is placed at the origin of coordinates, and its associated 
green and black bars describe adopted and formal uncertainties, respectively. The centre of 
the SPLE (mass distribution) is located at $\Delta x \sim$ 0\farcs6 from the origin, and its
associated green bars represent 1$\sigma$ errors.}
\label{fig:f13}
\end{figure} 

Using constraints from HST imaging, \citet{2019MNRAS.483.5649S,2021MNRAS.501.2833S} 
found a mass/light positional offset for G exceeding 0\farcs1 (see Fig. 5 in that paper). 
However, our solution from the SOC 2 suggests a positional offset of $\sim$0\farcs06 (see 
Figure~\ref{fig:f13}), which is one half of that by Shajib et al., but still noticeably 
large. This new offset is about 10 times the formal uncertainties $\sigma_x$ and $\sigma_y$ 
for G (see above). Shajib et al. have briefly commented on the existence of nearby 
companions of G, which could fix the issue. For example, \citet{2012A&A...538A..99S} 
reported that "astrometric anomalies" of many quads are solved by explicitly incorporating 
the nearest galaxy/group to the main lens into the lens model. The astrometric anomaly of 
\object{PS J0147+4630} might also be related to the substructure and azimuthal shape of the 
mass distribution of G \citep[e.g.][]{2012A&A...538A..99S,2021MNRAS.504.1340G}. 
Alternatively, the displacement vector from the observed centroid to the centre of the mass 
distribution could be real instead of an artefact due to the mass model used 
\citep[e.g.][]{2016ApJ...820...43S}. 

\begin{table}[h!]
\begin{center}
\caption{Predicted values of macrolens flux ratios and magnifications.}
\label{tab:t10}
\begin{tabular}{lc}
   \hline \hline
   Parameter & SPLE + ES (SOC 2)\\
   \hline 
   $M_{\rm{BA}}$ 		& 0.561 $\pm$ 0.001 \\
   $M_{\rm{CA}}$		& 0.567 $\pm$ 0.008 \\
   $M_{\rm{DA}}$    		& 0.056 $\pm$ 0.002 \\
   $\mu_{\rm{A}}$ 		& 36.4 $\pm$ 4.4    \\
   $\mu_{\rm{B}}$  		& 20.4 $\pm$ 2.4    \\
   $\mu_{\rm{C}}$  		& 20.6 $\pm$ 2.2    \\
   $\mu_{\rm{D}}$ 		& 2.03 $\pm$ 0.22   \\	
   \hline
\end{tabular}
\end{center}
\footnotesize{Note: Macrolens flux ratios $M_{\rm{XA}} = F_{\rm{X}}/F_{\rm{A}}$ (X = B, C, 
D) and macolens magnification ($\mu$) of the four quasar images.}
\end{table}

We may also consider the predicted macrolens flux ratios and magnifications in 
Table~\ref{tab:t10} as proxies for their true values. First, the $M_{\rm{BA}}$ and 
$M_{\rm{DA}}$ values in Table~\ref{tab:t10} are in good agreement with the $B/A$ and $D/A$ 
flux ratios for the [O\,{\sc iii}]$\lambda 5007$ emission line (see Table~\ref{tab:t5}). 
This narrow forbidden line is most likely related to a quasar extended region that is 
unaffected by microlensing \citep[e.g.][]{2002ApJ...576..640A}. Additionally, regarding the 
narrow-line emitting region in \object{PS J0147+4630}, differential dust extinction at 
$\lambda_{\rm{lens}} \sim$ 10\,000 \AA\ does not seem to play a relevant role in image pairs 
AB and AD\footnote{$\lambda_{\rm{lens}}$ denotes wavelength in the lens rest-frame}. The 
situation is quite different, however, for the pair AC. The C image is probably crossing a 
dust-rich region of the lens galaxy, which would resolve the conflict between $M_{\rm{CA}}$ 
and $C/A$ for the [O\,{\sc iii}]$\lambda 5007$ line. 

Second, if we focus on the images A and B, the $B/A$ flux ratio for the continuum at 
$\lambda_{\rm{rest}}$ = 5100 \AA\ (see Table~\ref{tab:t5}) is only 4$-$8\% above the value 
of $M_{\rm{BA}}$ in Table~\ref{tab:t10}. Therefore, $B/A$(cont@5100) is weakly affected by 
differential microlensing and differential dust extinction at $\lambda_{\rm{lens}}$ = 
10\,200 \AA\ (the time delay between A and B is extremely short). To account for the 4$-$8\%  
excess, the two simplest scenarios are: (1) A is not affected by extinction/microlensing, 
i.e. we can adopt a galaxy transmission factor $\epsilon_{\rm{G,A}}$ = 1 for light emission 
at 5100 \AA\ and a total magnification equal to $\mu_{\rm{A}}$ in Table~\ref{tab:t10}, but B 
is suffering a slight microlensing magnification ($\epsilon_{\rm{G,B}}$ = 1 and total 
magnification $\alpha \mu_{\rm{B}}$, where $\alpha$ = 1.06 $\pm$ 0.02 and $\mu_{\rm{B}}$ is 
given in Table~\ref{tab:t10}), and (2) B is not affected by extinction/microlensing 
($\epsilon_{\rm{G,B}}$ = 1 and total magnification $\mu_{\rm{B}}$), but the dust of G is 
weakly affecting A or microlensing is demagnifying this image ($\epsilon_{\rm{G,A}} \times$ 
total magnification = $\mu_{\rm{A}}/\alpha$). 

\section{Mass of the central black hole in the quasar}
\label{sec:bhmass}

Assuming the two extinction/microlensing scenarios in the last paragraph of Section 
\ref{sec:mass}, we plausibly estimated the mass of the supermassive black hole at the 
centre of \object{PS J0147+4630}. \citet{2006ApJ...641..689V} have reported a relationship 
between the central black hole mass in a quasar, the continuum luminosity at 
$\lambda_{\rm{rest}}$ = 5100 \AA, and the FWHM of the H$\beta$ emission line, and we used 
their result to obtain H$\beta$-based masses from the GTC-EMIR spectra of A and B in Section 
\ref{sec:spec}. While the H$\beta$ line widths, FWHM = 5731 (A) and 5588 (B) km s$^{-1}$, 
were used as they are, we adopted a flat $\Lambda$CDM cosmology with $H_0$ = 70 km 
s$^{-1}$ Mpc$^{-1}$ to infer luminosities from continuum fluxes $F_{\rm{5100~\AA}}$ = 30.21 
(A) and 17.92 (B) in units of 10$^{-17}$ erg cm$^{-2}$ s$^{-1}$ \AA$^{-1}$, the 
corresponding galaxy transmissions and total magnifications, and the Milky Way transmission 
$\epsilon_{\rm{MW}}$ = 0.95 \citep[see Eq. (1) in][]{2021A&A...646A.165S}. The first 
extinction/microlensing scenario led to ${\log \left[ M_{\rm{BH}}/\rm{M_{\odot}} \right]}$ =   
9.31$-$9.36 and 9.28$-$9.34 for A and B, respectively. The second scenario yielded ${\log 
\left[ M_{\rm{BH}}/\rm{M_{\odot}} \right]}$ = 9.32$-$9.37 (A) and 9.30$-$9.35 (B).
Additionally, using Eq. (4) of \citet{2011ApJ...742...93A} to estimate H$\beta$-based 
masses, the mass logarithm ranged from 9.30 to 9.40. 

\citet{2011ApJ...742...93A} also proposed a correlation between $M_{\rm{BH}}$, the continuum 
luminosity at $\lambda_{\rm{rest}}$ = 5100 \AA, and the FWHM of the H$\alpha$ emission line.
The line widths FWHM$_{\rm{H\alpha}}$ = 4981 (A) and 4943 (B) km s$^{-1}$ from the GTC-EMIR 
spectra, along with the continuum luminosities (see the previous paragraph) and Eq. (5) of 
\citet{2011ApJ...742...93A}, provided confirmation of the H$\beta$-based black hole masses. 
We obtained ${\log \left[ M_{\rm{BH}}/\rm{M_{\odot}} \right]}$ values ranging between  
9.29 and 9.37, in excellent agreement with those from the H$\beta$ line. Although our 
overall result is ${\log \left[ M_{\rm{BH}}/\rm{M_{\odot}} \right]}$ = 9.34 $\pm$ 
0.06, the scatter of $\pm$0.06 dex does not account for errors in line widths, continuum 
fluxes, and other physical quantities. Assef et al. showed (see their Table 5) that the true 
uncertainty in the logarithmic mass is four to five times larger than our scatter 
estimation. Thus, we finally adopted ${\log \left[ M_{\rm{BH}}/\rm{M_{\odot}} \right]}$ = 
9.34 $\pm$ 0.30. 

In a Schwarzschild geometry, the event horizon of the central black hole would be 
located at a typical radius of $6.5\times10^{14}$ cm, whereas the innermost ring of the 
accretion disc would have a radius three times larger, i.e. $\sim$$2\times10^{15}$ cm. 
Although the Event Horizon Telescope (EHT) has recently taken stunning radio images of the 
vicinity of the central supermassive black holes in the Milky Way and the local galaxy M87
\citep{2019ApJ...875L...1E,2022ApJ...930L..12A}, such tiny regions in distant galaxies 
cannot be resolved by direct imaging. Fortunately, inner regions of accretion discs in
gravitationally lensed quasars can be spatially resolved by microlensing 
\citep[e.g.][]{2004ApJ...605...58K,2021A&A...654A..70F} and reverberation-mapping 
\citep[e.g.][]{2012ApJ...744...47G} studies. Thus, the microlensing variability observed for 
14 lensed quasars provided a relationship between black hole mass and accretion disc radius 
at 2500 \AA\ \citep{2018ApJ...869..106M}, which yielded a radius of about $1.2\times10^{16}$ 
cm for the 2500 \AA\ continuum source of \object{PS J0147+4630}. Using the standard 
accretion disc model \citep{1973A&A....24..337S}, the measured black hole mass and predicted 
accretion disc size led to an Eddington factor ${\log \left( L/\eta L_{\rm E} \right)} \sim$  
2 \citep[e.g.][]{2010ApJ...712.1129M}, suggesting a very low radiative efficiency $\eta \sim 
0.01\left( L/L_{\rm E} \right)$.

\section{Conclusions}
\label{sec:end}

In this paper, we performed a comprehensive analysis of the optical variability of the 
quadruply-imaged quasar \object{PS J0147+4630}. Well-sampled light curves from its discovery 
in 2017 to 2021 were used to robustly measure the time delay between the brightest image A 
and the faintest D (167.5 $\pm$ 7.4 d, A is leading). Unfortunately, these light curves did
not allow us to accurately determine the very short time delays between the three bright 
images ABC forming a compact arc. Additionally, the A image was weakly affected by 
microlensing over the period 2017$-$2021, while the microlensing-induced variation of the C 
image was particularly large in that period. Combining our new brightness records with 
quasar fluxes from Pan-STARRS imaging in 2010$-$2013, the extended light curves also 
revealed significant long-term microlensing effects. A microlensing analysis of current data 
and future light curves from a planned optical multi-band monitoring is expected to lead to 
important constraints on the spatial structure of the quasar accretion disc 
\citep{2008A&A...490..933E,2008ApJ...673...34P,2020ApJ...905....7C,2020A&A...637A..89G}.

The GTC and Keck Observatory public archives contain unexplored near-IR spectroscopy of the
lensed quasar \object{PS J0147+4630} in 2018$-$2019. We extracted spectra for individual
images from such near-IR observations, and then analysed in detail the Mg\,{\sc ii}, 
H$\beta$, [O\,{\sc iii}], and H$\alpha$ emission lines, as well as their associated 
continua. These emission lines cover the spectral range 0.9$-$2.4 $\mu$m, are unaffected by 
absorption features, and were used to constrain the source (quasar) redshift in a reliable 
way. We obtained $z_{\rm{s}}$ = 2.357 $\pm$ 0.002, which resolves the controversy over the 
$z_{\rm{s}}$ value through visible spectra \citep{2017A&A...605L...8L,2018ApJ...859..146R}. 
We also derived single-epoch image flux ratios for emission lines, and the continuum at 
quasar rest-frame wavelengths of 2800, 5100 and 6563 \AA. Athough a detailed analysis of 
the flux ratios in Table~\ref{tab:t5} is out of the scope of this paper, it might provide 
accurate measurements of macrolens flux ratios, and reveal details of the spectral 
microlensing and dust extinction in the system \citep[e.g.][]{2016A&A...596A..77G,
2017ApJ...836...14S}. Spectral microlensing is often used to probe the quasar structure 
\citep[e.g.][]{2007A&A...468..885S,2012ApJ...755...82M,2021A&A...653A.109F}. In addition, 
we detected a very broad component in the H$\alpha$ emission that is likely related to the 
outflow in the BAL quasar \citep{2017A&A...605L...8L,2018ApJ...859..146R}.

Using HST imaging of the quad, \citet{2019MNRAS.483.5649S,2021MNRAS.501.2833S} have 
carried out reconstruction of the lensing mass from an SPLE + ES model. Adopting updated 
redshifts of the source and lens \citep[see here above and][]{2019ApJ...887..126G}, and 
assuming a standard cosmology with $H_0 \sim$ 70 km s$^{-1}$ Mpc$^{-1}$, the Shajib et al.'s 
solution cannot reproduce the measured delay between images A and D. An unacceptably high 
value of $H_0$ or an unusual external convergence is required to account for our longest 
delay. All the previous mass models, including the most recent modelling of 
\citet{2022arXiv220604696S}, actually predict a delay between the brightest image and the 
faintest that exceeds the measured one by a factor of $\sim$1.5$-$1.8. To take a deeper look 
at the SPLE + ES mass model, we used Gaia-HST astrometry and the measured delays as 
constraints. Updated redshifts and a standard cosmology with $H_0$ = 70 km s$^{-1}$ 
Mpc$^{-1}$ were also adopted. We found that the power-law index of the SPLE and the position 
angle of the ES disagree with those of the Shajib et al.'s solution. Although the Gaia-HST 
relative positions of quasar images and measured delays are well reproduced by the mass 
model, the light and mass distributions of the lens galaxy do not match. There is a 
significant mass/light misalignment that could be true or due to an oversimplification of 
the lens scenario \citep[e.g.][]{2012A&A...538A..99S,2016ApJ...820...43S,
2021MNRAS.504.1340G}. Further refinement of the new lens mass model along with an 
extension/improvement of the set of observational constraints (delays, macrolens flux 
ratios, galaxy velocity dispersion, etc) will contribute to an accurate determination of 
$H_0$ and other cosmological parameters \citep[e.g.][]{2017MNRAS.465.4914B,
2020A&A...643A.165B}. 

Comparing macrolens flux ratios predicted by our lens model with measured flux ratios 
(see above), we checked the consistency of results and discussed some 
extinction/microlensing effects in the lens system. The narrow-line emitting region is 
expected to be free from microlensing effects \citep[e.g.][]{2002ApJ...576..640A}, although 
it could be affected by dust extinction. Indeed, the flux ratios for the [O\,{\sc 
iii}]$\lambda 5007$ emission line are consistent with the absence of microlensing and the
presence of an important amount of dust in the region of the lens galaxy that is crossed by 
the C image. Additionally, the  $B/A$ flux ratio for the continuum at $\lambda_{\rm{rest}}$ 
= 5100 \AA\ is very close to the macrolens flux ratio between images B and A. Hence, with 
regard to the continuum at $\lambda_{\rm{rest}}$ = 5100 \AA, both images are presumably 
suffering weak extinction/microlensing effects, which allowed us to obtain a relatively 
narrow range of quasar luminosities from the fluxes of A and B. These luminosities, and the 
widths of the H$\beta$ and H$\alpha$ emission lines for the two brightest images, led to an 
estimate of the black hole mass in the heart of the distant quasar \citep[a black hole mass 
based on spectra of two quasar images in a quad was also derived by][]{2021A&A...656A.108M}. 
The mass of the black hole in \object{PS J0147+4630} is similar to those measured in other 
quads from Balmer line widths \citep[e.g. Cloverleaf quasar and Einstein 
Cross;][]{2011ApJ...742...93A}, but one of great relevance in constraining the relationship 
between accretion disc size and black hole mass at masses above $2\times10^{9}$ 
$\rm{M_{\odot}}$ \citep[see Fig. 9 of][]{2018ApJ...869..106M}.

\begin{acknowledgements}
We thank Martin Millon for making publicly available a Jupiter notebook that has greatly 
facilitated the use of the PyCS3 software. We also thank an anonymous referee for her/his 
comments and suggestions, which have helped us to improve the original manuscript. 
This paper is based on observations made with the 
Liverpool Telescope (LT) and the Nordic Optical Telescope (NOT). The LT is operated on the 
island of La Palma by Liverpool John Moores University in the Spanish Observatorio del Roque 
de los Muchachos of the Instituto de Astrof\'isica de Canarias with financial support from 
the UK Science and Technology Facilities Council. The NOT is operated by the Nordic Optical 
Telescope Scientific Association at the Observatorio del Roque de los Muchachos, La Palma, 
Spain, of the Instituto de Astrof\'isica de Canarias. The data presented here were in part 
obtained with ALFOSC, which is provided by the Instituto de Astrof\'isica de Andalucia (IAA) 
under a joint agreement with the University of Copenhagen and NOTSA. We thank the staff of 
both telescopes for a kind interaction. This work is also based on spectroscopic data from 
the GTC Public Archive at CAB (INTA-CSIC), as well as the Keck Observatory Archive (KOA), 
which is operated by the W. M. Keck Observatory and the NASA Exoplanet Science Institute 
(NExScI), under contract with the NASA. We have also used imaging data taken from the 
Pan-STARRS archive, the Barbara A. Mikulski archive for the NASA/ESA Hubble Space Telescope, 
and the archive of the ESA mission Gaia, and we are grateful to all institutions developing 
and funding such public databases. HD acknowledges support from the Research Council of 
Norway. This research has been supported by the grant PID2020-118990GB-I00 funded by 
MCIN/AEI/10.13039/501100011033. 
\end{acknowledgements}

\end{document}